%% file: main.tex
\documentclass[conference,compsoc]{IEEEtran}

\input{packages}

\input{macros}

\title{CaFA: Cost-aware, Feasible Attacks With Database Constraints \\ Against Neural Tabular Classifiers}

\begin{document}

\author{\IEEEauthorblockN{
    Matan Ben-Tov\IEEEauthorrefmark{1},
    Daniel Deutch\IEEEauthorrefmark{1},
    Nave Frost\IEEEauthorrefmark{2},
    and Mahmood Sharif\IEEEauthorrefmark{1}}
\IEEEauthorblockA{\IEEEauthorrefmark{1}Tel Aviv University
\IEEEauthorrefmark{2}eBay\\
Email: matanbentov@mail.tau.ac.il, \{danielde, mahmoods\}@tauex.tau.ac.il, nafrost@ebay.com}}

\maketitle

\input{abstract}

\input{intro}

\input{background}

\input{threat_model}

\input{tech_approach}
\input{exp-setup}

\input{exp-results}

\input{discussion}

\input{conclusion}

\input{acknowledgements}

\bibliographystyle{plain}
\bibliography{cited}

\appendices{}
\input{appendix}

\input{meta-review}

\end{document}

%% file: packages.tex
\usepackage{graphicx} 
\usepackage{xspace}

\usepackage{algorithm}
\usepackage{algpseudocode}
\usepackage{subcaption}

\usepackage{amssymb}  %
\usepackage{amsmath}  %
\usepackage{cancel}   %

\usepackage{svg}  %

\usepackage{graphicx}  %
\usepackage{multirow}  %
\usepackage{diagbox}   %
\usepackage{float}
\usepackage{makecell}  %

\usepackage{stix}  %

\usepackage[colorlinks,
        citecolor=blue,
        linkcolor=blue,
        urlcolor=black]{hyperref}

\usepackage{soul}

\usepackage{cite}

\usepackage{enumitem}

\usepackage{url}

\usepackage{booktabs}

\usepackage[leftcaption]{sidecap}

%% file: macros.tex
\newif\ifdraft
\draftfalse

\newif\ifpreprint
\preprinttrue
\ifdraft
  \newcommand{\todocolor}[1]{\textcolor{cyan}{#1}}
  \newcommand{\todocolorb}[1]{\textcolor{teal}{#1}}
  \newcommand{\todocolorc}[1]{\textcolor{red}{#1}}
  \newcommand{\todocolord}[1]{\textcolor{purple}{#1}}
  \newcommand{\todocolore}[1]{\textcolor{olive}{#1}}  
\else
  \newcommand{\todocolor}[1]{}
  \newcommand{\todocolorb}[1]{}
  \newcommand{\todocolorc}[1]{}
  \newcommand{\todocolord}[1]{}
  \newcommand{\todocolore}[1]{}
\fi

\sethlcolor{lime}
\newcommand{\markdiff}[1]{#1}

\newcommand{\figref}[1]{Fig.~\ref{#1}}

\newcommand{\tabref}[1]{Tab.~\ref{#1}}
\newcommand{\secref}[1]{\S\ref{#1}}
\newcommand{\secrefs}[2]{\S\ref{#1}--\ref{#2}}
\newcommand{\appref}[1]{App.~\ref{#1}}
\newcommand{\algoref}[1]{Alg.~\ref{#1}}

\newcommand{\ml}[0]{ML\xspace}
\newcommand{\cafa}[0]{CaFA\xspace}
\newcommand{\sys}[0]{\cafa{}\xspace}
\newcommand{\pgd}[0]{PGD\xspace}
\newcommand{\tabpgd}[0]{TabPGD\xspace}
\newcommand{\tabcw}[0]{CWL0\xspace}
\newcommand{\dcs}[0]{DCs\xspace}
\newcommand{\nn}[0]{NN\xspace}
\newcommand{\lpnorm}[1]{\ensuremath{\ell_{#1}}\xspace}

\makeatletter
\renewcommand{\ALG@beginalgorithmic}{\small}
\makeatother

%% file: abstract.tex
\begin{abstract}

This work presents \sys{}, a system for \textbf{C}ost-\textbf{a}ware
\textbf{F}easible \textbf{A}ttacks for assessing the robustness of
neural tabular classifiers against adversarial examples
realizable in the problem space, while
minimizing adversaries' effort. To this end,
\sys{} leverages \tabpgd{}---an algorithm we set forth to generate
adversarial perturbations suitable for tabular data---and incorporates
integrity constraints automatically mined by state-of-the-art
database methods. After producing adversarial examples in the
feature space via \tabpgd{}, \sys{} projects them on the mined
constraints,
leading, in turn, to better attack realizability.
We tested \sys{}
with three datasets and two architectures and found, among others,
that the constraints we use are of higher quality (measured via soundness and
completeness) than ones employed
in prior work. Moreover, \sys{} achieves higher feasible success
rates---i.e., it generates adversarial examples that are often
misclassified while satisfying constraints---than prior attacks
while simultaneously perturbing few features with
lower magnitudes, thus saving effort and improving inconspicuousness.
 \markdiff{We open-source \sys{},\footnote{\href{https://github.com/matanbt/attack-tabular}{https://github.com/matanbt/attack-tabular}}} hoping it will serve as a
generic system enabling machine-learning engineers to assess their
models' robustness against realizable attacks, thus advancing deployed
models' trustworthiness.

\end{abstract}

%% file: intro.tex
\section{Introduction}
\label{section:introduction}
\label{sec:intro}

Evasion attacks producing adversarial examples---slightly but
strategically manipulated variants of benign samples inducing
misclassifications---have emerged as a 
technically deep challenge posing risk to safety- and
security-critical deployments of machine learning (\ml{})~\cite{Biggio18Wild}. For
example, adversaries may inconspicuously manipulate their
appearance to circumvent face-recognition
systems~\cite{Cohen23NIR, mahmood-glasses, Sharif19AGNs}. As another example, 
attackers may introduce seemingly innocuous stickers to traffic signs,
leading traffic-sign recognition models to err~\cite{Evtimov18Traffic}. Such
adversarial examples have also become the {\it de facto} means for
assessing \ml{} models' robustness (i.e., ability to withstand
inference-time attacks) in adversarial
settings~\cite{Biggio18Wild, problem-space-terminology}. 
\markdiff{Nowadays, numerous critical applications employ \ml{} models
  on tabular data, including for medical diagnosis, malware detection,
  fraud detection, and credit
  scoring}~\cite{tabular-dl-survey-2021}\markdiff{. Still,
  adversarial examples against such models remain %
  underexplored.} 

\begin{figure}[t!]
\centerline{\includegraphics[width=\columnwidth]{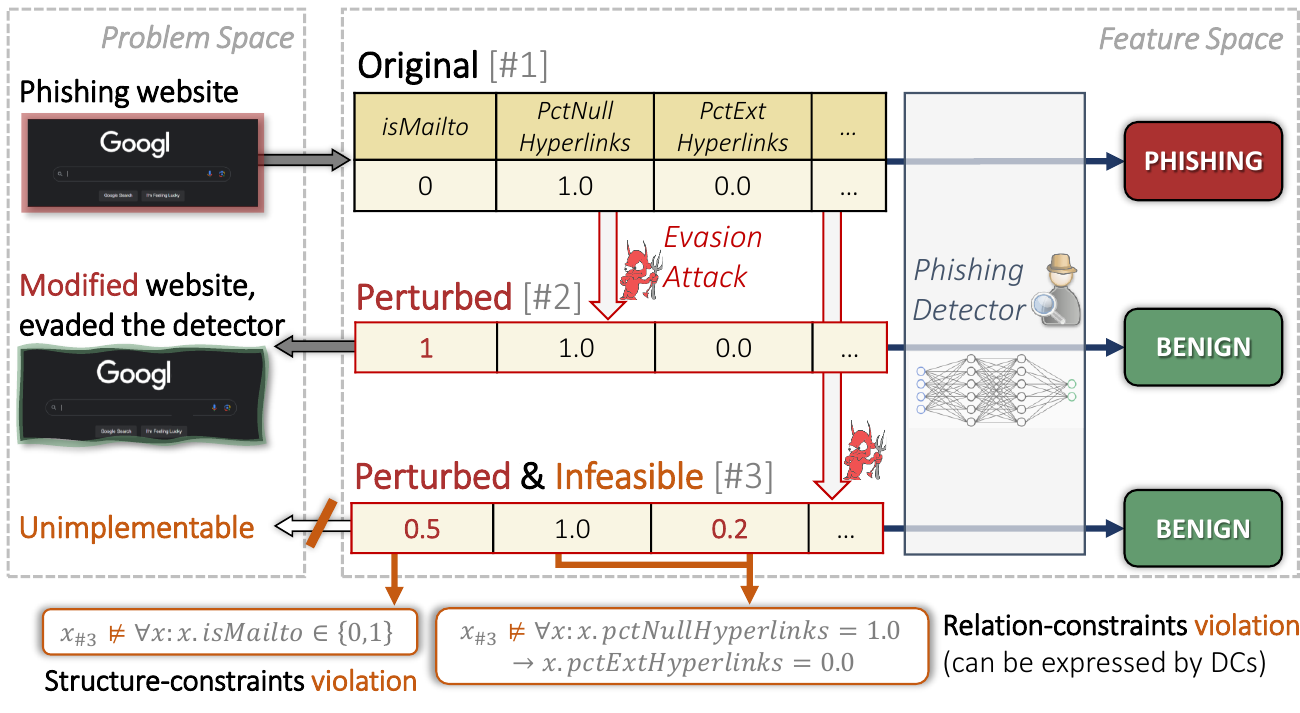}}
\caption{Adversarial examples in the problem space (e.g., a phishing
  website imitating Google) and the feature space (i.e., feature vectors
  serving as inputs to \ml{} models). The original feature
  vector (\textit{\textbf{\#1}}) represents a website correctly
  detected by \ml{}-based phishing detection. The adversary finds a
  minimal perturbation (\textit{\textbf{\#2}}) realizable as a
  problem-space instance while misleading the detector. Attacks,
  however, may also fail to satisfy data-integrity constraints
  (\textit{\textbf{\#3}}), rendering them unrealizable in the problem
  space. Different types of data-integrity constraints exist, including 
  structure (defined by features' domains) and relation (defined by
  relations between samples and features) constraints.
  } 
\label{figure:tabular-attack}
\end{figure}

In the tabular domain,
\ml{} models classify problem-space artifacts (e.g., phishing
pages or malicious programs) based on their feature representations
(e.g.,~\cite{problem-space-terminology,
  adult-dataset, bank-dataset, phishing-dataset}), posing
challenges to evaluating robustness against practical attacks
realizable in the problem space (i.e., via an artifact causing
misclassification). Some approaches for
evaluating robustness against practical attacks %
operate directly in the problem space by manipulating (realizable) 
artifacts directly to evade \ml{} models
(e.g.,~\cite{Evtimov18Traffic, mahmood-binary, mahmood-glasses,
  Sharif19AGNs, Song22MAB}). For
instance, some attacks suggest means to strategically transform
malicious programs such that their features would mislead malware
detectors~\cite{mahmood-binary, Song22MAB}.
These approaches, however, are domain-specific, rendering them
unsuitable for assessing robustness across multiple domains. Namely,
to evaluate %
robustness against practical attacks in domains not previously
studied, one still needs to invest manual effort and rely on domain
expertise to develop an attack for the specific
domain~\cite{Eykholt23URET, graphbased-kulynych-2019}.

In contrast, some techniques generate adversarial examples in the
feature space to assess model
robustness~\cite{problem-space-terminology}. These, 
however, often do not accurately reflect models' robustness in 
practice, as they produce feature changes not realizable in
the problem space due to violating data-integrity constraints~\cite{problem-space-terminology}
(\figref{figure:tabular-attack}).
While approaches for incorporating
constraints---e.g., in Valiant's boolean conjunctive normal
form (CNF)~\cite{valiants-paper-1984, sheatsley-constraints}---exist,
as we show (\secref{subsection:exp0-sound-vs-comp} and
\secref{subsection:exp-case-study}), such constraints may lack
soundness (i.e., adversarial examples satisfying the constraints often
violate realizability), 
and the attacks have limited success, leading to misapproximation
of true robustness.

Effort minimization poses another challenge for attacks on tabular
data. Specifically, adversaries typically seek to minimize adversarial
perturbations of original samples to maintain attack
inconspicuousness, thus generating harder-to-detect adversarial
examples (e.g.,~\cite{carmela-cost-utility, Grosse17AdvMal,
  szegedy-2014-intriguing, Sharif19AGNs}). Furthermore, 
as adversaries are typically rational and economically
motivated~\cite{Herley12Nigeria}, minimizing adversarial perturbations
can also help minimize attacks' (monetary)
costs~\cite{carmela-cost-utility}. Certain approaches incorporate
domain knowledge, assigning monetary value to perturbations of
different feature types~\cite{carmela-cost-utility}. These, however, cannot be
applied generically across domains. More generic approaches limit the
extent to which features can be perturbed (i.e., perturbations'
\lpnorm{\infty}- or \lpnorm{2}-norms) but may allow many features to be
manipulated and ignore the heterogeneity of features often encountered
in tabular data (e.g., binary vs.\ continuous features covering large
ranges)~\cite{simonetto-2022}. Other approaches limit the amount of
features that can be manipulated (i.e., perturbations' \lpnorm{0}-norms)
but may allow features to be manipulated to an arbitrary
extent~\cite{sheatsley-constraints, sheatsley-2022,
  android-bostani-2023, mathov-2021-surragate}. Thus, there remains a
need for approaches minimizing adversarial-example-generation effort in
a generic manner well-suited for tabular data.

To address these challenges, we propose \textbf{\sys{}}, a three-stage system
for \textbf{C}ost-\textbf{a}ware \textbf{F}easible
\textbf{A}ttacks to enable generic evaluation of \ml{} models
ingesting tabular data against evasion
(\secref{section:tech-approach}). \sys{} automatically
mines so-called denial constraints (DCs) via databases-based techniques~\cite{discovering-dc} identifying
relations between features and samples. Then, \sys{} perturbs samples
via \tabpgd{}---an algorithm we offer to generate evasive samples in
feature space while satisfying structure constraints defined by
feature developers and minimizing a newly introduced cost metric
accounting for the extent to which features are manipulated and the
number of features perturbed. Lastly, \sys{} 
projects the evasive samples on the mined DCs as a mean to
comply with genuine data-integrity constraints and ensure
realizability. We conducted 
experiments with three datasets and two neural network (\nn{})
architectures, exploring the quality of the mined constraints and the
feasible success rates (i.e., portion of adversarial samples
satisfying the mined constraints) attained by \sys{}. Our findings
show that:
\begin{itemize}[leftmargin=*]
   \item DCs better balance soundness (i.e., out-of-domain
    samples violate constraints) and completeness
    (i.e., in-domain samples satisfy constraints) than Valiant's
    constraints with tractable parameterization, rendering them more
    adequate for robustness evaluation 
    against practical attacks (\secref{subsection:exp0-sound-vs-comp}).
  \item \sys{}, with DCs employed, attains higher feasible success rates 
    (of over than $\times$25\%; \secref{subsection:exp1-real-success}), while perturbing
    relatively limited number of features to fewer extents
    (\secref{subsection:exp-2-cost}) than prior attacks. \sys{} also
    has markedly faster run times than attacks with
    competitive feasible success rates
    (\secref{subsection:exp-runtimes}). %
  \item When \dcs{} are {\it not} incorporated, \tabpgd{}
    with Valiant's constraints employed attained
    significantly higher feasible success rates than previously
    found (87.6\% vs.\ $\le$60.0\%), countering prior belief that
    integrating Valiant's integrity constraints harms
    attack success~\cite{sheatsley-constraints}
    (\secref{exp-valiants}).
     \item In a realistic phishing-page detection setting, adversarial
       examples satisfying the mined DCs found by 
    \sys{} could be realized in the problem space more successfully
    than ones found via prior techniques
    (\secref{subsection:exp-case-study}). %
    Intuitively, as the
    databases community pushes the boundaries of constraint mining,
    yielding constraints with enhanced completeness and
    soundness, \sys{}'s ability to assess robustness against practical
    attacks would also improve. 
\end{itemize}

Next, we discuss related work and background
(\secref{section:background}) and our 
threat model (\secref{section:threat-model}) before presenting our
technical approach (\secref{section:tech-approach}),
and results (\secrefs{section:expsetup}{sec:expresults}).
\markdiff{We wrap up with a discussion} (\secref{section:defenses})
and a conclusion (\secref{section:conclusion}).

%% file: background.tex
\section{Background and Related Work} \label{section:background}

\subsection{Evasion Attacks}
\label{sec:background:evasion}

Our work studies evasion attacks, popularized by the work of Biggio et
al.~\cite{biggio-2013} and Szegedy et
al.~\cite{szegedy-2014-intriguing}. In these attacks, adversaries
modify inputs at inference time, creating so-called adversarial
examples, to mislead \ml{} models.
Various methods with varied assumptions (e.g., full, white-box
vs.\ limited, query-only, black-box access) for generating adversarial
examples have been proposed~\cite{Croce20Auto, cw-paper, pgd-paper,
  papernot-2016-blackbox, jsma-paper, Biggio18Wild}. A common method
that serves as the bedstone for state-of-the-art attacks %
is Projected
Gradient Descent (\pgd{})~\cite{pgd-paper}. To produce adversarial
examples, \pgd{} iteratively modifies samples in the direction of the
gradient while limiting the magnitude of the perturbation. Concretely,
given a model with a loss function $L$ and a sample $x$ of class $y$,
\pgd{} iteratively calculates:
\[
x'^{(i+1)} := x'^{(i)} + \Pi (\alpha \cdot \text{sign} \\ (\nabla_x L(x'^{(i)}, y)))
\]
where $x'^{(i)}$ is the perturbed sample in the $i$-th iteration
($x'^{(0)}=x$), $\Pi$ is a function projecting samples per a
budget (e.g., clipping to an \lpnorm{\infty}-norm
$\epsilon$-ball centered at $x$), and $\alpha$ is a step size.
We propose a modification of \pgd{}
better suited for misleading models classifying tabular data
(\secref{subsection:tech-attack}).

\subsection{Challenges of Feature-Space Attacks}
\label{subsection:background-realistic-feature-space-attacks}

To evade \ml{} models in practice, adversaries need to modify
problem-space artifacts to induce misclassifications (e.g., altering phishing
websites to evade detection)~\cite{problem-space-terminology}. To the
best of our knowledge, there exists no technique capable of
generating problem-space attacks in a generic way, across
domains.
Consequently, domain-specific attacks for generating
adversarial examples in the problem space are constantly being
proposed, even when attacks in closely related domains already
exist. Indeed, such attacks have been proposed for domains ranging
from malicious PDF detection~\cite{Xu16PDF} to %
malware detection~\cite{Song22MAB}, and from phishing
detection~\cite{Panum20Phish}   
to traffic-sign recognition~\cite{Evtimov18Traffic}.

As a generic alternative to problem-space attacks, one may generate
adversarial examples in the feature space that can be later mapped to
problem-space instances. Such an approach faces two primary challenges.
First, feature-space attacks are often \textit{unrealizable} in the
problem space~\cite{problem-space-terminology}. This is often due to
violating data-integrity constraints not accounted for during adversarial
example generation (e.g., see example in \secref{subsection:background-constraints}).
While specific attempts to produce realizable adversarial examples in
the feature space exist (e.g.,~\cite{Eykholt23URET, Grosse17AdvMal}), they
usually rely on domain expertise to define permissible perturbation
ensuring the evasive features are realizable, rendering them non-generic.
Moreover, as we find (\secref{subsection:exp1-real-success}), the
heterogeneous features with varying scales and 
types (e.g., categorical and continuous) hinder the success of
standard attacks used in prior work.
Second, feature-space attacks typically require high
\textit{adversarial effort}, either
due to manipulating many features (mainly, when minimizing
\lpnorm{2}- or \lpnorm{\infty}-norms)~\cite{simonetto-2022}, or due to manipulating a few
features by significant amounts (when minimizing
\lpnorm{0}-norm)~\cite{sheatsley-constraints, sheatsley-2022,
  android-bostani-2023, mathov-2021-surragate}.
This, in turn, renders adversarial examples easy to
detect~\cite{Evtimov18Traffic, Grosse17AdvMal, Sharif19AGNs}, and
may increase the monetary cost of crafting them
in the problem space~\cite{carmela-cost-utility}. As mentioned above
(\secref{sec:intro}), this work seeks to overcome these challenges.

\subsection{Data-Integrity Constraints}
\label{subsection:background-constraints}

Identifying and accounting for feature-space constraints
in attacks can help improve realizability.
In our work, we capture feature-space constraints with the known structural limitations of the feature space, and with the integrity
constraints learned from the dataset itself, both are then integrated
into our attack.
Further, we evaluate and compare integrity-constraint sets by measuring the
extent of violations detected (i.e. soundness) and the extent of
in-domain samples admitted (i.e. completeness) (see
\secref{subsection:metrics}). We focus on two constraint
types---structure and relation constraints.

\subsubsection{Structure Constraints}
Non-realizability sometimes stems from violations of basic
feature-space structural properties. For instance, in our example
(\figref{figure:tabular-attack}), $x_3.\mathit{isMailto}=0.5$, while
$\mathit{isMailto}$ is a binary feature, stating whether the HTML uses the
\textit{mailto} function or not. Thus, it is unclear how $x_3$
can be transformed, if at all, to an evasive HTML in the
problem space. To address these kind of violations, we
specify the type of each feature, which can be
\textit{ordinal}, \textit{continuous}, or
\textit{categorical}. For each feature, we then define a permissible
domain (i.e., range or set of values). Additional
structure constraints may stem from the feature processing (e.g., from
one-hot encoding of categorical features). Such structure constraints
can be typically derived from the mapping defined during feature
engineering.

\subsubsection{Relation Constraints}
Non-realizability may also arise from
violations of semantic relations between features. For instance,
in our example ($x_3$ in \figref{figure:tabular-attack}), while the
proportion of empty links ($\mathit{pctNullHyperlinks}$) is 100\%, the
proportion of external links ($\mathit{pctExtHyperlinks}$) is
20\%, which is impossible for real-world HTMLs.
Preventing this type of complex cross-feature violations requires
expressive modeling. We employ
the highly expressive Denial Constraints
(\dcs{})~\cite{discovering-dc} to capture relation constraints. We
also refer to and compare with Valiant's
constraints~\cite{valiants-paper-1984}.

\textbf{Valiant's Constraints.} Valiant's PAC learning
algorithm~\cite{valiants-paper-1984} derives a set of boolean
constraints, as CNFs, from a given dataset. The algorithm and its
constraints were previously used to capture domain constraints by
Sheatsley et al.~\cite{sheatsley-constraints}. In this work, we employ
Valiant's constraints as a baseline for feature-space constraints (\secref{subsection:exp-setup-constraints}).

\textbf{Denial Constraints.} \dcs{} are widely used and
provably expressive type of data-integrity constraints (e.g., \dcs{}
can express the most commonly used constraint
types~\cite{discovering-dc, adcminer-paper}).
\dcs{} apply to
pairs of samples in the dataset and are composed of negated
conjunctions of predicates. Namely, a DC expresses a set of predicates
that \textit{cannot} hold together for a pair of samples. The set
of possible \dcs{} can be formalized as: 
\begin{align*}
&\left\{ \forall x,x'\in X.\lnot\left(\bigwedge_{p\in P}p\right)\mid P\subset\Psi\right\}
\\ &\quad \text{s.t.} \quad \Psi:=\left\{ \left(x_{i}\diamond x'_{j}\right)\mid\diamond\in\{=,\neq,<,>,\leq,\geq\},i,j\in\left[d\right]\right\} 
\end{align*}
where $\Psi$ is the set of possible predicates, which can vary in
different flavors of \dcs{}. Each predicate includes a \textit{main}
tuple, $x$, the \textit{other} tuple $x'$, and an operator between
these tuples, $\diamond$. The set of possible operators may naturally
differ between features (e.g., categorical features can only use the
operators ${=, \neq}$). Using this definition, \dcs{} can also express
the relation constraint violated in \figref{figure:tabular-attack}
by instantiating the \textit{other} tuple values as constant
operands. Due to their expressivity, we employ DCs as to capture
feature-space relation constraints.

Approaches to \textit{mining} constraints from a given dataset
can be roughly divided into two. The first mines \textit{hard
  constraints}, which requires the mined constraints to hold for
\textit{all} samples in the learned set. 
The other, more expressive type of constraints, namely, \textit{soft
  constraints}, allows small violation rates in the learned set. 
In contrast to other constraints mining algorithms---such as
Valiant's algorithm~\cite{valiants-paper-1984}---that mine hard
constraints, we leverage algorithms for mining soft DCs due to their
lower sensitivity to malformed or anomalous samples~\cite{fastadc,
  adcminer-paper}.

\subsection{Existing Attacks on Tabular Datasets}
\label{sec:back:PriorTabAttacks}

Adversarial examples in the tabular domain have gained
an increasing attention in recent years. Still, these
remain relatively under-explored compared to attacks in the vision and
text domains. \tabref{tab:relworkcomp} summarizes and compares prior work
and ours, focusing on the aforementioned challenges.

\input{tables/related-works-comparison}

\textbf{Realizability Through Constraints.}
Previous works on tabular attacks acknowledge the importance of
adhering to constraints imposed by the problem-space when crafting
adversarial samples~\cite{sheatsley-constraints, android-bostani-2023,
  sheatsley-2022, mathov-2021-surragate}.  Relatedly, an analogous
claim was made for finding better counter-factual examples by Deutch
and Frost~\cite{daniel-constraints}. Most prior attacks have focused
on structure constraints, whose significance was also discussed by
Mathov et al.~\cite{mathov-2021-surragate}. 

While many efforts acknowledge relation constraints,
only a few suggested automatic and general schemes utilizing
them (\tabref{tab:relworkcomp}). 
Notably, Sheatsley et al.'s closely related
work~\cite{sheatsley-constraints} uses constraints mined by Valiant's
algorithm~\cite{valiants-paper-1984}. They incorporate the constraints
into their attacks, for which they report low rates of misclassified
adversarial samples, thus arguing that data-integrity constraints
increase robustness. We later 
show findings contradicting Sheatsley et al.'s
(\secref{exp-valiants}). Moreover, we argue that \dcs{} are better
aligned with the genuine problem-space constraints relative to
Valiant's constraints (\secref{subsection:exp-case-study}).

There are several ways to integrate feature-space
constraints into attacks. Simonetto et al.~\cite{simonetto-2022}
minimize a penalty associated with the constraints to improve
adversarial examples' compliance.
Another option is applying
SAT solvers to project adversarial examples onto the constrained
feature space~\cite{sheatsley-constraints, simonetto-2022}; however,
this approach demonstrated poor feasible success
rates~\cite{sheatsley-constraints}. In this work, we also leverage SAT
solvers, but precede projection with an attack well-suited for tabular data
(\secref{subsection:tech-attack}) and
employ new heuristics during projection to preserve misclassification
while simultaneously satisfying constraints
(\secref{subsection:tech-projection}). Our experiments demonstrate 
that, prior to projection on constraints using solvers, our approach
already produces fewer violations of relation constraints than past
attacks, lending itself to higher feasible success rates after
projection (\secref{subsection:exp1-real-success}).

\textbf{Adversarial Effort (Cost).} The literature varies in its
definition of \textit{cost} for tabular adversarial example attacks. Some
efforts adopt metrics commonly used in computer vision
(\lpnorm{2}- or \lpnorm{\infty}-norms)~\cite{simonetto-2022}; some
argued for the suitability of the \lpnorm{0} cost for tabular
data~\cite{sheatsley-2022, sheatsley-constraints,
  mathov-2021-surragate}; while others formed novel cost
measures (e.g., feature-importance-based
cost~\cite{important-features-perturb, zoo-on-fraud-sony}). Keeriv 
et al.~\cite{carmela-cost-utility} offered a nuanced discussion on the
importance of cost in tabular attacks, distinguishing it from
the imperceptibility goal in the vision domain and describing it as
the financial cost of the attack. %
Accordingly, to bound costs, they require manually defined financial
costs for each feature. Their interpretation and incorporation of 
cost in the attack is conceptually close to ours, albeit our
attack's cost is automatically derived. 

Overall, as can be seen from \tabref{tab:relworkcomp}, our proposed
system, \sys{}, is the first that assesses robustness to realizable
attacks by incorporating constraints while simultaneously minimizing
adversarial effort automatically and generically across domains,
regardless of feature types. As our experiments show
(\secrefs{section:expsetup}{sec:expresults}),
\sys{} also attains higher attack success-rates at lower costs than
leading attacks, providing more reliable robustness estimation.

%% file: tables/related-works-comparison.tex
\begin{table*}[hbt!]
    \centering
  \resizebox{\textwidth}{!}{%
  \begin{tabular}{l| l | c c c| l c | c | c}
  \hline
  
  \hline
  
 \multirow{2}{*}{\bf \diagbox[width=8em]{Work}{Criteria}}   & \multirow{2}{*}{\textbf{Setting}} & \multicolumn{3}{c|}{\textbf{Constraints}}& \multicolumn{2}{c|}{\textbf{Cost}}& \multirow{2}{*}{\textbf{Domain-General?}}& \multirow{2}{*}{\textbf{Open Source?}} \\

    & %

    & \textbf{Structural?}
    & \textbf{Relational?}
    & \textbf{Automated Mining?}

    & \textbf{Type}
    & \textbf{Automated?}

    & 
    \\ \midrule

    Ballet et al.~\cite{important-features-perturb} & White-box \
    &$\dottedcircle$& $\dottedcircle$& - & Feature importance & $\mdlgblkcircle$ & $\mdlgblkcircle$ &   $\mdlgblkcircle$  \\  \midrule

    Bostani et al.~\cite{android-bostani-2023} & White-box& $\mdlgblkcircle$ & \makecell{$\mdlgblkcircle$ \\ (via statistical correlations)} %
    & \makecell{$\mdlgblkcircle$ \\ (OPF algorithm)} &$\ell_0$& $\mdlgblkcircle$ &\makecell{$\dottedcircle$\\(only binary features)} &  $\mdlgblkcircle$\\ \midrule
    
    Cartella et al.~\cite{zoo-on-fraud-sony} & Black-box 
    &$\circlebottomhalfblack$& $\dottedcircle$& - &Feature importance & $\mdlgblkcircle$ & $\mdlgblkcircle$ & $\dottedcircle$\\ \midrule

    Keeriv et al.~\cite{carmela-cost-utility} & Black-box 
    & $\mdlgblkcircle$ & $\dottedcircle$& - & Monetary cost &$\dottedcircle$& $\mdlgblkcircle$ & $\dottedcircle$ \\ \midrule
    
    Mathov et al.~\cite{mathov-2021-surragate} & Black-box &
    $\mdlgblkcircle$ & \makecell{$\circlebottomhalfblack$\\ (via trained encoder model)}
    & $\mdlgblkcircle$ &
    $\ell_0$ &$\mdlgblkcircle$ & $\mdlgblkcircle$ & $\dottedcircle$ \\ \midrule

    Sheatsley et al.~\cite{sheatsley-constraints} & White-box & $\mdlgblkcircle$
    & \makecell{$\mdlgblkcircle$ \\ (Valiant's)} & \makecell{$\mdlgblkcircle$ \\(Valiant's)} &$\ell_0$ & $\mdlgblkcircle$ & $\mdlgblkcircle$ & $\dottedcircle$ \\ \midrule
    
    Sheatsley et al.~\cite{sheatsley-2022} & White-box 
    & $\mdlgblkcircle$ &  $\mdlgblkcircle$ &
    \makecell{$\circlebottomhalfblack$ \\ (manual \& heuristics)} &
    $\ell_0$ & $\mdlgblkcircle$ & $\mdlgblkcircle$ & $\dottedcircle$ \\ \midrule
    
    Simonetto et al.~\cite{simonetto-2022} & White-box (C-PGD)
    & $\mdlgblkcircle$ &  $\mdlgblkcircle$ & $\dottedcircle$ &$\ell_\infty / \ell_2$ & $\mdlgblkcircle$ & $\mdlgblkcircle$ & $\mdlgblkcircle$ \\ \midrule
    
    Simonetto et al.~\cite{simonetto-2022} & Grey-box (MoEvA2) 
    & $\mdlgblkcircle$ &  $\mdlgblkcircle$ & $\dottedcircle$ &$\ell_\infty / \ell_2$ & $\mdlgblkcircle$ & $\mdlgblkcircle$ & $\mdlgblkcircle$ \\ \midrule

    \sys{} (ours) & White-box & $\mdlgblkcircle$ & \makecell{$\mdlgblkcircle$ \\ (DCs)} & \makecell{$\mdlgblkcircle$ \\(FastADC~\cite{fastadc})} &\textit{standardized-}$\ell_\infty$  $+$  $\ell_0$& $\mdlgblkcircle$ & $\mdlgblkcircle$ & $\mdlgblkcircle$ \\

    \bottomrule
  \end{tabular}}
    \caption{\label{tab:relworkcomp}Comparing prior attacks on
      tabular data in terms of incorporating constraints for
      realizability, attack-cost minimization, generality across
      feature domains, and open-source availability. Each
      cell is marked to denote whether attacks support
      ($\mdlgblkcircle$), partially support
      ($\circlebottomhalfblack$), or do not support
      ($\dottedcircle$) certain properties.
    }
\end{table*}

%% file: threat_model.tex
\section{Threat Model} \label{section:threat-model}

In this paper, we study white-box feature-space
attacks incorporating learned constraints to produce feasible (in
terms of the chosen constraint set) and cost-aware adversarial 
examples. We focus on targeting Neural Network (\nn{}) models, due to
their popularity~\cite{sheatsley-constraints},
performance~\cite{tabnet-2020}, and other desirable properties (e.g.,
lending themselves to distributed
training and self-supervised learning)~\cite{tabular-dl-survey-2021}.

\begin{SCfigure*}[][t!]
\centering
\includegraphics[width=1.3\columnwidth]{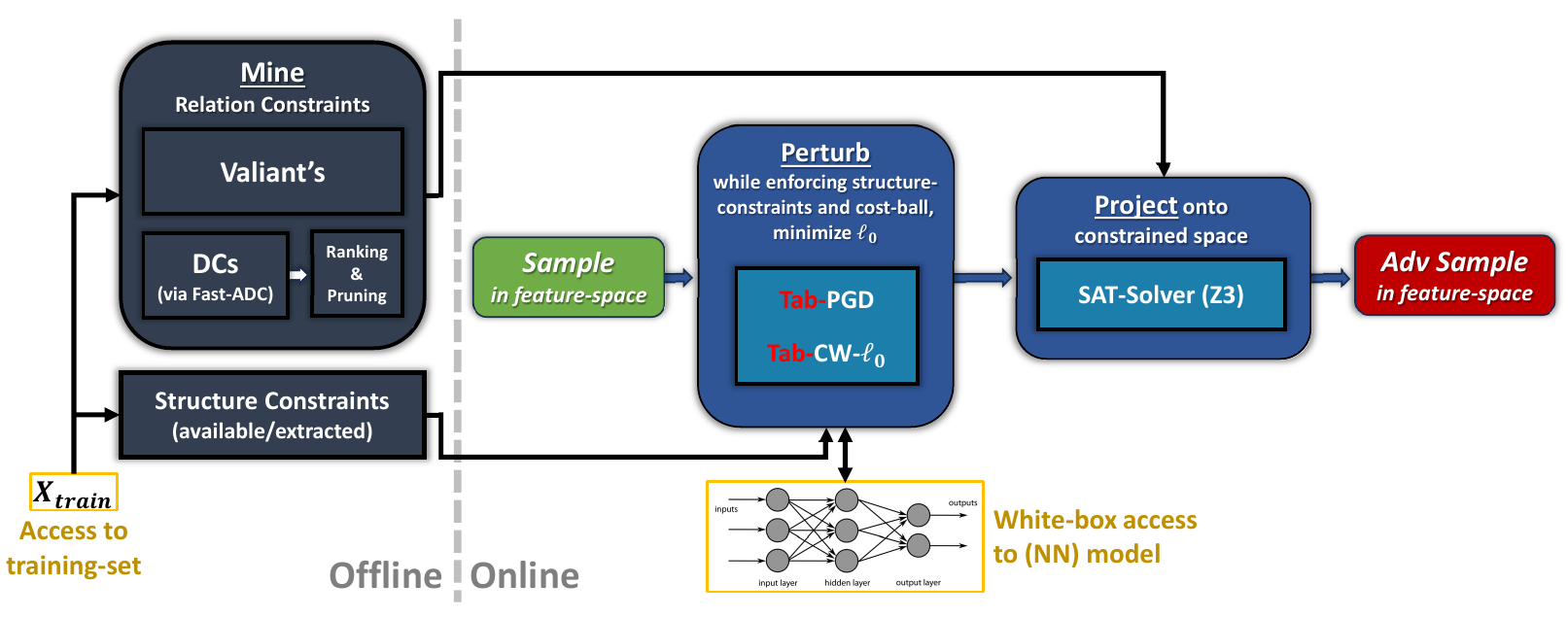}
\caption{\label{figure:attack-flow}\sys{}'s full flow, including the
  \textit{offline} phase---where \sys{} mines feature-space constraints
  (\secref{subsection:mining})--and the \textit{online}
  phase--where \sys{} incorporates constraints when generating adversarial examples.
  In the online stage, we initially find adversarial perturbations
  that both comply with the structure constraints
  and mislead the model (\secref{subsection:tech-tabpgd-attack}). Then,
  we project the adversarial sample to comply with the 
  mined relations constraints (e.g., \dcs{})
  (\secref{subsection:tech-projection}).}
\end{SCfigure*}

We study adversaries targeting \nn{}-based tabular data classifiers,
denoted by \(M:X\to Y\), where \(X 
\subseteq \mathbb{R}^d\) is the feature space
and \(Y\) is the label space. \(X\) is composed of heterogeneous
features, each from a different 
(possibly dependent) unknown distribution.  
We assume adversaries possess a %
sample $x\in X$ they seek to
misclassify and a corresponding label $y\in Y$.

We consider a white-box setting where the attacker
knows the classifier's architecture. %
The attacker also has access to the inherent structure
constraints of $X$ defined during feature engineering, as well as to
a sample dataset for mining relation
constraints. %
These assumptions, considering the worst-case adversary,
are standard and common in the literature
(e.g.,~\cite{Baluja18ATNs, Carlini22MemInf, membership-inference-paper}).

The adversary's primary goal, given $x$ and $y$, is to craft a
realistic feature-space instance misleading the \ml{} model. Such instance
can be found through a feature-space attack, with the following
objective:
\begin{align*}    
&\text{Find}\quad \delta\in \mathbb{R}^d \quad \\ 
&\text{s.t.}\quad 
M\left(x+\delta\right)\neq y,\; (x+\delta)\models T,\; \\
& \quad \mathit{Cost}(\delta)\leq B
\end{align*}
Namely, the adversary aspires to find a feature-space perturbation
($\delta$) that modifies the targeted sample ($x$) to be
\emph{misclassified} by the model;
the modified sample must be \emph{feasible} under the feature-space
constraints ($T$), lending the sample to better problem-space
realizability; and
the modification of the original problem-space sample should involve
\emph{minimal costs} (i.e., adversarial effort), which is embodied by a
computable $\mathit{Cost}$ function and a budget bound ($B$) in the
feature space.

%% file: tech_approach.tex
\section{Technical Approach}
\label{section:tech-approach}

To address the challenges set forth in the previous sections, we propose \sys{}, 
a \textit{feature-space attack} targeting \nn{} models, which crafts \textit{cost}-aware
adversarial examples that are feasible relative to
\textit{feature-space constraints}.
  As depicted in
\figref{figure:attack-flow}, we divide our method to three
stages, similarly to Sheatsley et al.~\cite{sheatsley-constraints}:
(1) mining the feature constraints and defining their utilization (\secref{subsection:mining});
(2) perturbing the given sample to fool the classifier model while
maintaining \textit{structure constraints} and low cost
with \tabpgd{}-\tabcw{} (\secref{subsection:tech-tabpgd-attack});
and (3) projecting the perturbed sample onto the learned constrained space
using the Z3 solver~\cite{z3-solver-paper} (\secref{subsection:tech-projection}).

\subsection{Mining Constraints} \label{subsection:mining}
As previously mentioned, we seek constraints of two
types---\emph{structure constraints} and \emph{relation constraints}
in the form of DCs.
While the former are usually readily available (as defined during
feature engineering), those of the latter type are typically not given
explicitly and require extraction from the data using specialized
mining frameworks.

To mine DCs, we use FastADC \cite{fastadc}, a
state-of-the-art technique for discovering \textit{soft} \dcs{}. We
opt for soft constraints as hard constraints can limit and potentially
exclude valuable and insightful constraints due to anomalous
samples (e.g., a key constraint or a trend valid for $\ge$99\%
of samples might be omitted because of a malformed sample or rare
feature value). We mine the constraints from the training set,
tolerating up to a predefined violation rate among the dataset's
sample pairs. %

When using \dcs{} in the attack, we seek to perturb samples while
complying with \dcs{}. Recall that a DC accounts for a \textit{pair}
of samples---\textit{other} and \textit{main}
(\secref{subsection:background-constraints}). Accordingly, to test
the compliance of a sample with a DC, we are required to assign the
perturbed sample as the \textit{main} sample and assign each of the
training samples $\in{}X_\mathit{train}$ as the \textit{other} sample, resulting in
$|X_\mathit{train}|$ \textit{practical} constraints per DC. Thus, for a large set of
DCs, $D$, the number of \textit{practical} constraints becomes excessively large
($|D|\cdot{}|X_{train}|$). Such a large number of constraints may harm
run-time performance, thus necessitating filtering the DC sets. This
filtration also allows us to control the constraints'
soundness-completeness trade-off (see \secref{subsection:metrics} and
\secref{subsection:exp0-sound-vs-comp}). Namely, removing low-quality
DCs covering a few samples can increase completeness while incurring limited
decrease in soundness.

To select well-performing \dcs{} during filtration, we form a
ranking-scheme based on well-established metrics~\cite{fastadc,
  discovering-dc} quantifying succinctness 
(favoring concise constraints), coverage (favoring constraints
more supported by data), and certain forms of constraint violations
(see \appref{appendix-dcs-ranking}). Subsequently, we pick the
top-ranked constraints and closely related tuples. More specifically,
we pick the highest-ranked \dcs{} according to the metrics (amount
denoted by $n_\mathit{dcs}$), and, for each DC, we pick the
\textit{other} tuples that provided the best compliance with the DC
(an amount denoted by $n_\mathit{tuples}$, uniform over all \dcs{}).
This process allows evaluating multiple DC sets and choosing a well-sized,
high-quality set of constraints, which we denote as $T$.

\subsection{Perturbing Samples}
\label{subsection:tech-attack}\label{subsection:tech-tabpgd-attack}
  
Now we present \tabpgd{} and \tabpgd{}+\tabcw{}, novel evasion attacks
for crafting adversarial samples complying with
\textit{structure constraints}, and minimizing adversaries' effort.
We mathematically define the cost function
(\secref{subsubection:tech-attack-cost}), refer to the employed
structure constraints
(\secref{subsubection:tech-attack-structure-constraints}), and,
finally, fully describe the \tabpgd{} attack
(\secref{subsubection:tech-attack-algo-tabpgd}) and its \tabcw{}
extension (\secref{subsubection:tech-attack-algo-cw}). The samples
crafted with this attack are later projected onto relation constraints
(\secref{subsection:tech-projection}).

\subsubsection{Heterogeneous Cost} \label{subsubection:tech-attack-cost}

As a proxy to the adversarial effort, our
attack aspires to bound and minimize the adversarial perturbation with 
a $\ell_\infty$-norm variant (\textit{standardized-}$\ell_\infty$) and
the $\ell_0$-norm, respectively. Specifically, in our framework, we
bound the first (\tabpgd{}) then minimize the second
(\tabcw{}). In doing so, we seek to make attacks more inconspicuous
and, by proxy, limit their attacks' financial cost. Incidentally, as a byproduct, we empirically find that this approach also increases
compliance with relation constraints, possibly due to limiting
deviation from original samples that already satisfy these constraints
(\secref{subsection:exp1-real-success}), \markdiff{and helps preserve
  functionality in the problem space}
(\secref{subsection:exp-case-study}).

\textbf{\textit{standardized-}$\ell_\infty$-norm} ensures small-magnitude changes in ordered features (i.e., continuous and ordinal).
To account for the heterogeneity of tabular data, we perform
a min-max scaling before calculating the $\ell_\infty$-norm. Said
differently, given a range size of each feature, \(R_i\), derived from the
\(i\)\textsuperscript{th} feature's support, we standardize the features before
applying the $\ell_\infty$-norm. Formally, 
\(\text{\textit{standardized-}}\ell_\infty(\delta) = \max_{i}
\left|\frac{\delta_i}{R_i}\right|\), where $i$ iterates over the
ordered features.
We define the \textit{standardized-}$\ell_\infty$ $\epsilon$-ball
accordingly:
$\forall i: [x_i - \epsilon \cdot R_i, x_i + \epsilon \cdot R_i]$,
and call the $R_i$s as the standardization factors.

\textbf{$\ell_0$-norm} minimization limits the overall number of
features that are altered.  

Since each tabular feature mostly refers to a different subdomain, one
can interpret the first cost as minimizing the \textit{extent} of
required effort within each subdomain, and the second
cost as minimizing the \textit{variety} of efforts required. Both goals are 
essential to ensure minimal effort when transforming feature-space
samples to the problem space.

\subsubsection{Structure Constraints} \label{subsubection:tech-attack-structure-constraints}
Our attack incorporates the targeted dataset's structure constraints
(\secref{subsection:background-constraints}), utilizing each feature's
\textit{type} and \textit{set of permissible values}. Furthermore, we
account for the \textit{encoding methods} of categorical
features. Chiefly, for MLPs
(\secref{subsecion:exp-setup-targeted-models}), our attack seeks to
preserve one-hot encodings' syntax, and, for TabNets, it handles
multi-dimensional continuous embeddings of discrete features
 (\secref{subsection:exp1-tabnet}). For both encoding methods, our
attack follows the same principled approach
(\secref{subsubection:tech-attack-algo-tabpgd}).

\nn{}s often utilize one-hot encodings to process categorical
features~\cite{categorical-encoding-paper}, thus capturing features'
categories without imposing arbitrary order on feature
values. Specifically, given a categorical feature \(f\) with \(S\) categories, its one-hot encoding is a binary
vector \(OneHot(f) \in \{0,1\}^{|S|}\), where the \(j\)\textsuperscript{th} entry is 1 if
\(j=f\) and 0 otherwise. Consequently, preserving valid encoding
introduces an additional \textit{structure constraint} on the feature
vector:
(1) each encoded coordinate is a single binary feature; and
(2) exactly one coordinate in \(OneHot(f)\) must be 1.

\subsubsection{\tabpgd{}'s Algorithm}
\label{subsubection:tech-attack-algo-tabpgd}

\tabpgd{} tailors the \pgd{} framework to tabular data. 
\markdiff{\pgd{}, originally developed for the vision domain, ignores
  structure constraints and uses a fixed step size for all features.
To overcome limitations,}
\tabpgd{} involves \textit{heterogeneous} update steps and a cost bound ($\textit{standardized-}\ell_\infty$), and
incorporates structure constraints throughout the perturbation. A
concise pseudocode is provided in \algoref{alg:short-tabpgd} and
explained in what follows.

\begin{algorithm}[t!]
 \caption{\tabpgd{}}
 \label{alg:short-tabpgd}
\begin{algorithmic}[1]
\Require{$x$}$\text{ (sample)}$, {$y$}$\text{ (label)} $,

{$ \epsilon$}$\text{ (\textit{standardized-}$\ell_{\infty}$ bound)}, $

{$ \alpha$} $\text{(step-size factor)}, R \text{ (standardization factor)} ,$

{$ n$}$\text{ (iterations)}  $, \textit{structureConstraints}
\Ensure $x'$ is adversarial example, or $\bot$ if the attack fails.

\State $x' \gets \text{randomInitUnderStructureConstraint}(x) $
 \State $g^{(accum)} \gets 0 $

\For{$\_ \in \{0, 1,2,\dots, n\}$}
    \State $g \gets \nabla_{x'}  \text{CrossEntropy}(M(x'), y) $
    \State  $x'_{temp} \gets x' + \alpha \cdot (R \odot sign(g))$
    \State $g^{(accum)} \gets g^{(accum)} + g_{[\text{categorical coordinates}]}$  

    \State $x'\gets \text{perturbContinuousFeatures}(x'_{temp}, x')$
    \State $x'\gets \text{perturbIntegerFeatures}(\lceil x'_{temp}\rceil, x')$
    \State $x'\gets \text{perturbCategoricalFeatures}(g^{(accum)}, x')$
    
    \State $x' \gets \text{clipToRange}(x')$
    \State $x' \gets \text{clipTo $\epsilon$ Cost}(x')$

    \If {$M(x') \neq y$}
        \State \textbf{return} $x'$
    \EndIf 
    
\EndFor
\State \textbf{return} $\bot$
\end{algorithmic}
\end{algorithm}

\tabpgd{} starts (Line 1) by randomly initializing the adversarial
sample ($x'$) within the \textit{standardized-}$\ell_\infty$
$\epsilon$-ball around the original sample
($x$). It then performs $n$ iterations, each updating $x'$ using
gradients.  Similar to \pgd{}, we calculate loss gradients w.r.t to
the perturbed sample ($g$, Line 4), then, we set the temporary
perturbation ($x'_\mathit{temp}$, Line 5) using the gradient's sign, a
predefined step size ($\alpha$), and the standardization
factor ($R$) (to account for heterogeneity, unlike PGD).
Since the temporal perturbation ($x'_\mathit{temp}$) may violate
structure constraints (specified by \textit{StructureConstraints}), we
update the perturbations to comply with these constraints:
\emph{continuous features} are left unchanged (Line 7); \emph{ordinal,
integer features} are rounded to the closest integer (Line 8); and
\emph{categorical features} (Line 9) are modified based on the
accumulated gradient vector ($g^{(accum)}$) of their encoding (e.g.,
one-hot encoding), selecting the category encoded by the largest accumulated gradient (across iterations) while
preserving valid encoding.
To further enforce \emph{structure constraints} and the \emph{cost
bound} we keep the perturbed features within their allowed ranges
(Line 10) and within the \textit{standardized-}$\ell_\infty$
$\epsilon$-ball (Line 11).
The algorithm terminates once the perturbed sample fools the model
(Lines 12-14) or after $n$ iterations without success (Line 16).

\subsubsection{\tabcw{} Variant}
\label{subsubection:tech-attack-algo-cw}
While \tabpgd{} aims to bound the
\textit{standardized-}$\ell_{\infty}$
cost of ordered features, it may still alter many features, thus
increasing adversaries' effort. To address this, \tabcw{} augments
attacks to account for the overall $\ell_0$ cost. This extension
adapts Carlini and Wagner (CW)'s \lpnorm{0}-norm
attack~\cite{cw-paper}\markdiff{---shown effective in reducing the} \lpnorm{0}-\markdiff{norm of perturbations in gradient-based attacks---}to work in tandem with \tabpgd{}.

\tabcw{} operates by running \textit{\tabpgd{}} iteratively as a
black-box. Each \tabpgd{} run generates a perturbation, $\delta$, and
its corresponding loss gradient w.r.t to the perturbed sample, $g$.  
We employ a modified version of CW's heuristic, assigning each feature $i$
a score
\[
    p_i(\delta, g, R) = \frac{g_i \cdot \delta_i}{R_i}
\]
to rank its importance for the attack; intuitively, the higher $p_i$ the more essential the feature for the attack. Note that, if the
$i$\textsuperscript{th} feature is a coordinate in a one-hot encoding,
we sum the importance over all of the encoding's coordinates, to form
a unified score for the categorical feature. After each iteration, the
feature with the lowest importance, $\text{argmin}_i\{p_i\}$, is
frozen, meaning it remains unperturbed in subsequent iterations. 
The algorithm iterates, freezing one feature at a time, until either
it fails to make further improvements or reaches a predefined maximum
number of iterations.

\subsection{Projecting on Relation Constraints}  \label{subsection:tech-projection}

Equipped with mined constraints ($T$, \secref{subsection:mining}),
\sys{} eventually
projects the preliminary adversarial sample, which, at this
point, complies with simple \textit{structure constraints}, onto
the constrained space imposed by the more complex mined
\textit{relation constraints} (e.g., DCs).
\markdiff{We employ SAT Solvers to guarantee the obtained samples
  satisfy the constraints after projection}, in a manner similar to
prior work~\cite{sheatsley-constraints}, \markdiff{in contrast to
  approaches treating the constraints as approximated
  objectives}~\cite{simonetto-2022}.
We implement the projection as follows:
\sys{} generates a first-order logic formula, $\Phi_T$, that captures $T$---the set of mined constraints---and uses a SAT solver to
ensure compliance. Furthermore, we introduce $\phi_T$ with additional
assertions to
verify the perturbations comply with structure constraints and
adhere to attack-defined cost metrics (by bounding the allowed projection).

When a perturbed sample, $x'$,
violates  $\Phi_T$, we attempt to satisfy it with partial assignment
by relaxing a portion of the literals (previously termed the
\textit{projection budget}~\cite{sheatsley-constraints}).  
Utilizing Sheatsley et al.'s heuristic~\cite{sheatsley-constraints},
we prioritize relaxing the features that are least constrained.  
Namely, we relax the features satisfying the fewest assertions independently
(intuitively, avoiding the most constrained features would increase
the odds of a successful projection). 
We further optimize this by employing binary search on the required
budget, aiming to minimize the number of relaxed features while
ensuring a successful projection. 
We empirically found both methods (ranking and binary search) helpful
for keeping the number of projected features minimal, thus helping to
preserve attack success after projection.

Having decided which literals to
free, our problem is reduced to solving partially-assigned $\Phi_T$,
with the partial assignments given by the non-relaxed
  features of $x'$. To do so, we employ the \textit{Z3}
Solver~\cite{z3-solver-paper}, obtaining the
projected, perturbed sample satisfying both structural and
relation constraints.

%% file: exp-setup.tex
\section{Experimental Setup}
\label{section:experiments}
\label{section:expsetup}

Next, we elaborate on the evaluation setup. We define the metrics we
used (\secref{subsection:metrics}), describe the datasets
(\secref{subsection:exp-setup-datasets}) and \nn{} models
(\secref{subsecion:exp-setup-targeted-models}), specify the
constraints employed (\secref{subsection:exp-setup-constraints}), and
list the evaluated attacks (\secref{subsection:exp-setup-attacks}). 

\subsection{Metrics} \label{subsection:metrics}

We used various metrics to evaluate constraints' quality and
attack efficacy.

\subsubsection{Constraint Metrics}
To evaluate constraints, we used and extended well-established metrics
previously used in the space~\cite{sheatsley-constraints}.

\textbf{Compliance} checks if a sample \( x \) satisfies all
constraints in \( T \):
    \[
    \forall t\in T: x \models t  \quad \text{, or simply:}  \quad x\models T.
    \]
If true, we say $x$ complies with $T$.

\textbf{Completeness} checks if \textit{all} feature-space samples
\( \in{}X \) comply with \( T \): 
    \[
        \forall x \in \mathbb{R}^d : \quad x \in X \implies x \models T.
    \]
If this holds, we say $T$ is complete relative to the feature space $X$.
    
\textbf{Soundness} checks if all instances (real vectors) complying
 with \( T \) pertain to the genuine feature space (i.e., \( \in{}X \)):
    \[
    \forall x \in \mathbb{R}^d : \quad x \models T \implies x \in X.
    \]
If they do, we say $T$ is sound relative to $X$.  

Since the mathematical definitions of soundness and completeness are not
feasible to compute (they require enumerating all samples in an
unknown feature space), we define empirical counterparts, lending
themselves to being computed on given a test set
$X_\mathit{test} \subset X$ and a set of constraints to evaluate $T$.

\textbf{Empirical Completeness} computes the proportion of \(
X_\mathit{test} \) samples complying with \( T \).
\[
\widehat{Completeness}_T(X_{test}) = \frac{\left|\left\{ x\in X_{test}\mid x\models T\right\} \right|}{\left|X_{test}\right|}
\]

\textbf{Empirical Soundness}, a metric we introduce,
leverages manually constructed, ground-truth
constraints (called golden constraints; denoted as
\(\mathring{T}\))~\cite{adcminer-paper} to test whether the mined
constraint set can detect violations of data-integrity constraints
known to hold.
For each sample \(x \in X_{test}\) that complies with our learned
constraints \(T\), we generate a \textit{modified} sample \(\mathring{x}\)
by intentionally violating a golden constraint
\(\mathring{t}\in\mathring{T}\). The violating \textit{modification} is done by
taking the negation of the constraint and uniformly sampling a random
feature
value that necessarily satisfies this negation. This creates
out-of-domain samples that \emph{should} also violate \(T\) if it is
sound. %

Hence, empirical soundness quantifies the proportion of modified
samples ($\mathring{x}$) that, while complying with $T$ \textit{prior}
to the modification, violate $T$ \textit{after} the modification. We
apply this prior restriction to samples that originally comply with
$T$, since we aspire to attribute the violations exclusively to
violating golden constraints. We consider each golden constraint
separately in the metric, as their mining complexity may vary:
    \begin{multline*}
        \widehat{Soundness}_T(X_{test},\mathring{T}) = \\ 
        \frac{\left\{ x\in X_{test},\mathring{t}\in\mathring{T}\mid x\models T\land \mathring{x}\cancel{\models}T\right\} }{\left|\mathring{T}\right|\cdot\left|\left\{ x\in X_{test}\mid x\models T\right\} \right|}
    \end{multline*}

$\textbf{F1}$ derives a composite score from (empirical) soundness and
completeness. Attaining perfect completeness (by accepting all
samples) or soundness (by rejecting all constraints) alone is trivial,
hence there is a need to balance both. One can view completeness and
soundness as analogous to recall and precision~\cite{Davis06PR},
respectively. Thus, following this analogy, we adopt the F1
score---the harmonic mean of completeness and soundness---as a quality
measure of the constraints set, seeking to maximize it. Formally:
\[
  F1 := \frac{2\cdot completeness \cdot soundness}{ completeness +
    soundness}.
\]
    
\subsubsection{Attack Metrics}
We use a battery of metrics to estimate attacks' efficacy.    

\textbf{Attack Success-Rate} measures the proportion of misclassified adversarial samples:
\[
     \frac{|\{ M(x+ \delta_x ) \neq  y \mid x \in X_{test} \}|} {| X_{test} |}
\]
where $ \delta_x $ is the adversarial perturbation the attack yields
for $x$. This is a widely accepted metric for evaluating attacks
(e.g.,~\cite{carlini2019evaluating, mahmood-glasses}).

\textbf{Feasible Attack Success-Rate} computes the proportion of
misclassified samples that also comply with \( T \).
\[
     \frac{|\{ M(x+ \delta_x ) \neq  y \ \land \ (x+\delta_x) \models T  \mid x \in X_{test} \}|} {| X_{test} |}  
\]
Related work also used this metric (e.g.,~\cite{simonetto-2022,
  sheatsley-constraints}).

\textbf{Cost} is measured in \(\ell_0\) and
\textit{standardized-}\(\ell_\infty\) (\secref{subsection:tech-tabpgd-attack}).

\subsection{Datasets} \label{subsection:exp-setup-datasets}
We used three commonly used~\cite{daniel-constraints,
  sheatsley-constraints, tabular-dl-survey-2021} tabular datasets.
All datasets were randomly split into 
training and testing sets (87\% and 13\% ratio). %
We summarize the features of each dataset in
\tabref{table:datasets-sum}.

\textbf{Phishing.} The \textit{phishing}
dataset~\cite{phishing-dataset}, for detecting phishing websites,
comprises 10K samples: 5K phishing and 5K legitimate websites. It
includes features derived from webpage URLs and HTML source code. We
focus on 10 out of 48 features that achieve maximal
model accuracy, as per work proposing the set~\cite{phishing-dataset}
and Sheatsley et al.~\cite{sheatsley-constraints}. %

\textbf{Bank Marketing.} This \textit{bank}
dataset~\cite{bank-dataset}, for predicting whether clients
would purchase bank services, has 45K samples, with 5.2K positive
labels indicating service purchase. We selected 11 out of 16 features
leading to the highest validation accuracy, as per Borisov et al.'s survey \cite{tabular-dl-survey-2021}.

\textbf{Adult.} The \textit{adult}
dataset~\cite{adult-dataset}, for predicting whether people's income
levels surpasses a certain threshold, contains 32.5K samples, with 8K
classified as high-income. We used all 13 available
features.

\begin{table}[t!]
  \resizebox{\columnwidth}{!}{%
  \begin{tabular}{lr|rrrr}
  \toprule
  & & \multicolumn{4}{c}{\bf \# Features by type} \\
  {\bf Dataset} & {\bf \# Samples} &  {\bf Categorical} & {\bf Continuous} & {\bf Ordinal} & {\bf All} \\
  \midrule  
    Phishing & 10K & 5 &  2 & 3 & 10   \\
    Bank &  45K    & 4 &  2 & 5 & 11   \\
    Adult &  32.5K   & 7 &  0 & 6 & 13 \\
  \bottomrule
  \end{tabular}}
  \caption{\label{table:datasets-sum}Dataset sizes and feature counts by feature types.}
\end{table}

\subsection{\ml{} Models} \label{subsecion:exp-setup-targeted-models}

We tested two \nn{} architectures, training models for each
dataset. First, we used MLPs with ReLU
activations, as is standard~\cite{tabular-dl-survey-2021,
  sheatsley-constraints}. For these models, we preprocessed our data
with standard techniques~\cite{tabular-dl-survey-2021}, applying one-hot encodings
for categorical features and coordinate-wise normalization based on
training-data statistics. The MLPs'
hyperparameters were determined through a grid search, optimizing for
benign accuracy (i.e., accuracy on clean, unperturbed data).
The chosen models have five hidden layers, each with width of 128,
and were trained using the Adam 
optimizer~\cite{adam-optimizer-paper} with a learning rate of
$5\times 10^{-4}$. Besides targeting MLPs, we experimented
with a transformer-based state-of-the-art architecture geared for
tabular data, TabNet~\cite{tabnet-2020},
choosing its hyperparameters via a grid search, and using %
the official implementation.
\tabref{table:arch-and-acc} reports the models' benign accuracy on the
three datasets. For all datasets, the models achieved accuracy
comparable to or exceeding past
benchmarks~\cite{tabular-dl-survey-2021, phishing-dataset,
  bank-dataset}. The TabNet models achieved slightly better, yet
similar, benign accuracy to their MLP counterparts.

\begin{table}[t!]
  \centering
    \begin{tabular}{l|rrr}
      \toprule

      {\bf \diagbox[width=8em]{Model}{Dataset}} & {\bf Adult} & {\bf Bank} & {\bf Phishing} \\ \midrule
      {\bf MLP} & 86.1\% & 89.0\% & 94.7\%  \\
      {\bf TabNet} & 87.0\% & 89.7\% & 95.9\%  \\
      
    \bottomrule
  \end{tabular}
  \caption{\label{table:arch-and-acc}The benign accuracy achieved over the test sets.}
\end{table}

\subsection{Constraints} \label{subsection:exp-setup-constraints}
We evaluated two types of relation constraints and used them to
measure feasibility. Particularly, we used DCs to assess feasibility,
except for \secref{exp-valiants}, where we used Valiant's.

\textbf{Valiant's Constraints.} We used a variant of Valiant's
algorithm, proposed by Sheatsley et al.~\cite{sheatsley-constraints},
to mine Valiant's Constraints in CNF form. 
Motivated to model the most general constraint theory, Sheatsley et al.\ proved
that their parameterization mines the least constrained constraint theory.
\markdiff{We note that any other parameterization of Valiant's is
  practically intractable to run, due to the exponential complexity of
  its mining algorithm
  (\appref{subsubsection:mining-valiants})}.

\textbf{Denial Constraints (DCs).} We mine \dcs{} with the
state-of-the-art FastADC algorithm~\cite{fastadc}, allowing violation
rates of up to 1\% for each soft constraint (i.e., no more than 1\%
of training samples can violate each constraint), similarly to
Livshits at al.~\cite{adcminer-paper}. 
(We clarify that this is different from the completeness metric from
\secref{subsection:metrics}, estimating the rate at which \emph{test
samples} violate \emph{all} constraints.) For performance reasons
imposed by FastADC, we restrict the set of possible DC predicates to
only compare between the same features. Note that cross-feature
comparisons are mostly irrelevant due to semantic differences between
features in tabular data. Yet, relations between features can still
be learned, as the constraint itself considers multiple features
(each in its own predicate). Additionally, we identified the predicate
space as a significant bottleneck in FastADC's runtime, 
stemming from an exponential dependency of the algorithm in the predicate space's
size.
Thus, formally, %
we used the predicate space:
\[
\Psi:=\left\{ \left(x_{i}\diamond x'_{i}\right)\mid\diamond\in\{=,\neq,<,>,\leq,\geq\},i\in\left[d\right]\right\}.
\]
This formalism can still capture expressive constraints of the form
presented in \figref{figure:tabular-attack} (and does so in practice).
The specific set of \dcs{} utilized in experiments was
chosen after evaluating multiple sets (see
\secref{subsection:exp0-sound-vs-comp}).

\subsection{Evaluated Attacks}
\label{subsection:exp-setup-attacks}
Besides the attack variants we propose, we evaluated and compared
multiple attacks whose hyperparameters were selected as to optimize
the feasible attack success-rates. \markdiff{Prior attacks were chosen
  for close characteristic with
  \cafa{}}~\cite{sheatsley-constraints,simonetto-2022}\markdiff{;
  other attacks were either difficult to properly reproduce in the
  absence of an implementation, did not involve relation constraints,
  or were domain specific }(see \tabref{tab:relworkcomp}).
    
\textbf{\sys{}} (\secref{section:tech-approach}) is our proposed
framework composed of several components. We set
\textbf{\tabpgd{}}'s %
step size $\alpha$ to $\frac{\epsilon}{100}$, its
\textit{standardized}-\lpnorm{\infty}'s
$\epsilon$ to $\frac{1}{30}$ (a choice which we further explore
in \secref{subsection:exp-2-cost}) and the number of iteration to 100.
\tabpgd{} %
can be augmented with \tabcw{} and followed by projection with a SAT solver.
We ran \tabcw{} for up to 30 iterations and used the Z3
solver~\cite{z3-solver-paper}.
Projections were done by freeing a subset of the features, ranging from 0\% to 50\% of
features and determined by binary search.

\textbf{PGD}~\cite{pgd-paper} is  an iterative gradient-based
white-box attack, limiting perturbations' $\ell_{\infty}$-norms
(\secref{sec:background:evasion}).
We set the norm bound ($\epsilon$) to 100 and the step-size ($\alpha$) to
1, ran attacks 100 iterations with early stopping, and used the
cross-entropy loss.
    
\textbf{C-PGD}~\cite{simonetto-2022} is a \pgd{} variant
transforming  constraints into a penalty function that is added
to the attack's loss function (\secref{sec:back:PriorTabAttacks}). The
attack optimizes three objectives---penalty function (constraints), misclassification,
and distance. We adopted the same parameters as in \pgd{}, assigning 0.1
as the weight for the additional penalty, and used the official
implementation.
      
 \textbf{MoEvA2} \cite{simonetto-2022} is a query-based attack using
 models' output probabilities and a multi-objective genetic algorithm
 (R-NSGA-III~\cite{nsga3})
 to optimize an analogous loss to C-PGD's. We used the
 official implementation and adopted the variant bounding the
 $\ell_2$-norm due to the higher feasible attack success rates
 attained compared to the \lpnorm{\infty}-norm variant (see
 \appref{appendix:moeva-norm-choice}). 
 The remaining parameters were adopted from the original work and
 implementation.

\textbf{Sheatsley et al.'s Attacks}\cite{sheatsley-constraints}
\markdiff{include (1) a PGD variant with structural constraints,
  and (2) a Constrained-Saliency Projection (CSP) attack, an
  $\ell_0$-norm-based attack}~\cite{jsma-paper},
\markdiff{both integrating Valiant's constraints. We compared these
  attacks to \sys{} in a 
  setting identical to the one considered in the original work (i.e.,
  under Valiant's constraint space).}

%% file: exp-results.tex
\section{Experimental Results}
\label{sec:expresults}

Now we turn to our results. We start with evaluating different
constraint sets' quality (\secref{subsection:exp0-sound-vs-comp})
before assessing attacks' effectiveness in terms of feasible success
rates under \dcs{} (\secref{subsection:exp1-real-success}). Then, we
test attacks on two other dimensions---attack cost
(\secref{subsection:exp-2-cost}) and run time 
(\secref{subsection:exp-runtimes}). Subsequently, we measure attacks'
feasible success rates with Valiant's constraints
(\secref{exp-valiants}). We close the section with a case study of
real-world phishing websites exploring the extent to which \sys{} and
other attacks can produce adversarial samples implementable in the
problem space (\secref{subsection:exp-case-study}).
        
\subsection{Constraints' Quality} \label{subsection:exp0-sound-vs-comp}

\textbf{Experiment Description.}
We empirically evaluated the completeness and soundness
(\secref{subsection:metrics}) of 
relation-constraint sets of different types to identify high-quality
ones. Specifically, we evaluated constraints mined with Valiant's
algorithm~\cite{valiants-paper-1984, sheatsley-constraints} 
and DCs mined with FastADC~\cite{fastadc}. Based on this, we found
constraints that best balance completeness and soundness, maximizing
the $F1$ score (\secref{subsection:metrics}). Here, we report on the
evaluation using the test set of the \textit{bank} dataset. We further
validate the findings on the \textit{phishing} dataset in
\appref{exp0-sound-vs-comp-phishing}. 

For this experiment, we used a set of 1.5K samples data points sampled from the \textit{bank} dataset's test set, where we mined the constraints from the
disjoint training set. %
For measuring soundness, we used the four golden constraints
identified by Deutch and Frost for the the \textit{bank}
dataset~\cite{daniel-constraints}; \markdiff{notably, such constraints
  are not readily available for other datasets, where a manual effort
  was required to extract them} (see
\appref{exp0-sound-vs-comp-phishing}). We list these golden
constraints in 
\tabref{table:bank-golden-constraints} (e.g., the first constraint 
requires that for a client, $x$, who was not contacted in the previous
marketing campaign (i.e., $previous$=0), the outcome of the previous
campaign is of category \textit{unknown} (i.e., $poutcome$=-1)). We
used these constraints to generate violating samples
(\secref{subsection:metrics}) that should be rejected by the mined
constraints.

\input{tables/bank-golden-constraints.tex}

We evaluate multiple sets of DCs, each of different size, to attain
different soundness-completeness tradeoffs: increasing the number of
constraints should increase soundness, whereas decreasing it should
boost completeness. We formed the subsets by picking the highest
ranked constraints, per the ranking
scheme previously presented (\secref{subsection:mining}), limiting
both the required amount of constraints (\textit{$n_{dcs}$}) and the
\textit{other} tuples ($n_{tuples}$).  

\begin{figure}[t!]
\begin{center}
\centering
\centerline{\includegraphics[width=0.9\columnwidth]{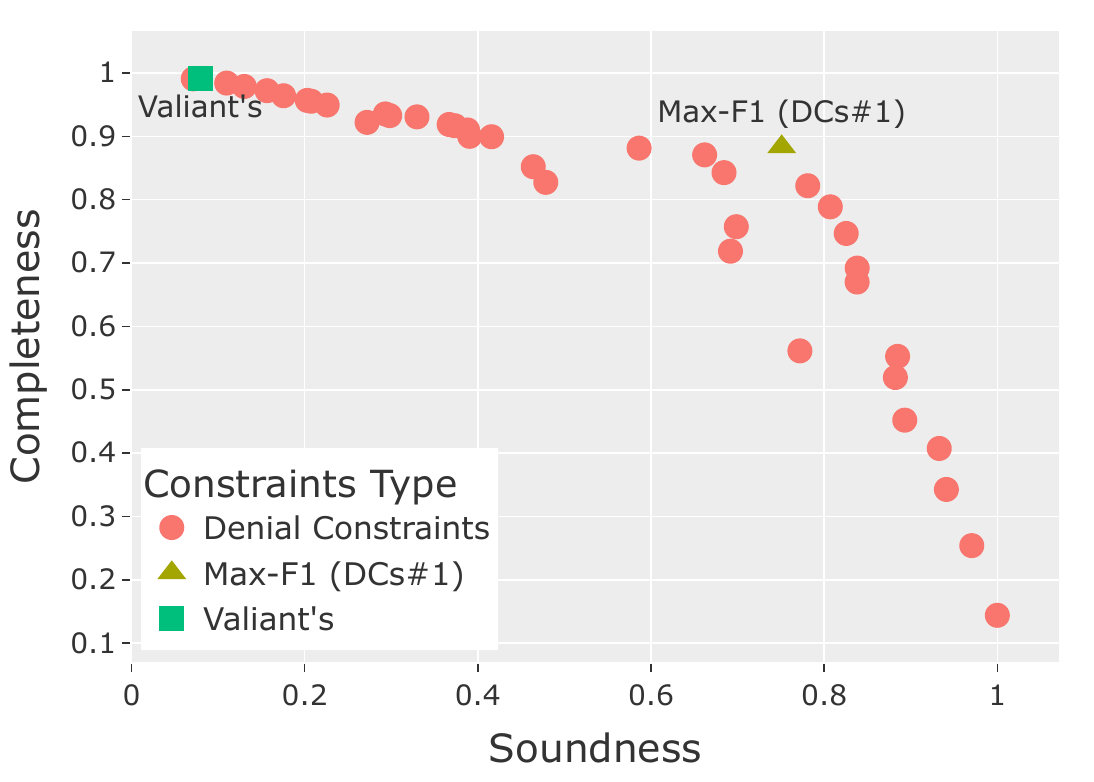}}
\caption{Evaluation of constraints' soundness and completeness
  over the test set. %
  We distinguish Valiant's constraints set from the rest of the \dcs{},
  mined by FastADC~\cite{fastadc}. We also mark \dcs{} set with the
  highest F1 score that we use in most experiments.} 
\label{figure:sound-vs-comp-picks} 
\end{center}
\end{figure}

\input{tables/exp0-comp-vs-sound-picks}

\textbf{Experiment Results.}
The empirical completeness and soundness for different constraint sets
are depicted in
\figref{figure:sound-vs-comp-picks}. \tabref{table:com-sound-picks} 
presents constraint sets of interest.
From our evaluation, we conclude that Valiant's constraints serve as a
strong baseline in terms of completeness---consistent with the inherent
properties of their mining process
(\secref{subsection:exp-setup-constraints})~\cite{sheatsley-constraints}.
To achieve this performance, Valiant's algorithm crafts
numerous constraints (51K; see \tabref{table:com-sound-picks}). In
contrast, \dcs{} manage to achieve comparable soundness and
completeness measures with significantly fewer constraints
(0.1K; \dcs{} \#3 in \tabref{table:com-sound-picks}). 
Furthermore, Valiant's algorithm yields low
\textit{soundness}.
\markdiff{This can be attributed to the algorithm configuration, which
  favors completeness over soundness. We reiterate that other
  parameterizations are infeasible to run
}(\secref{subsection:exp-setup-constraints})\markdiff{, preventing us
  from attaining other soundness-completeness trade-offs with
  Valiant's constraints.}
  Note that, since
soundness is monotone in the constraint sets (i.e., adding more
constraints to a set implies, by definition, larger soundness),
inspecting subsets of Valiant's full constraint set (as done for DCs) could only harm soundness.

The expected completeness-soundness trade-off is visible in the
analysis of \dcs{}---larger sets lead to higher soundness and lower
completeness (see \textit{\dcs{} \#2} and \textit{\dcs{} \#3} in 
\tabref{table:com-sound-picks}).
To balance the two metrics, picking performant \dcs{} for the
remaining experiments, we chose the constraint set achieving highest
F1 measure (\figref{figure:sound-vs-comp-picks}), resulting in a set
containing 5K constraints (\textit{\dcs{} \#1},
\tabref{table:com-sound-picks}). Applying the same evaluation on the
\textit{phishing} dataset, we reached similar conclusion
(\appref{exp0-sound-vs-comp-phishing}), leading us to use the 5K
constraints as a default configuration for \dcs{}. The chosen \dcs{}
sets of \textit{phishing}, \textit{adult}, \textit{bank} datasets
achieved 95.4\%,  84.7\%, and 88.3\%
empirical completeness, respectively, on the test data.

\subsection{Feasible Attack Success} \label{subsection:exp1-real-success}

We evaluate attacks' success, in terms of misleading the targeted
models (MLPs and TabNets) and feasibility of the adversarial
samples relative to the chosen set of \dcs{}.
In addition to evaluating \textit{full} attacks, when \dcs{} were integrated, we also
examined variations where we prevented access to \dcs{} by the attacks 
(e.g., by disabling projection after \tabpgd{} in \sys{}, or not
including constraints in C-PGD's and MoEvA2's losses). Doing so
helped us gain further insights about how attacks work and allowed us to estimate success when adversaries have no access to auxiliary data
for mining constraints.

\subsubsection{Attacking MLPs} \label{subsection:exp1-real-success-mlps}
\begin{figure*}[bht]
\begin{center}
\begin{subfigure}[t]{1.0\columnwidth}
    \centerline{\includegraphics[width=0.95\columnwidth]{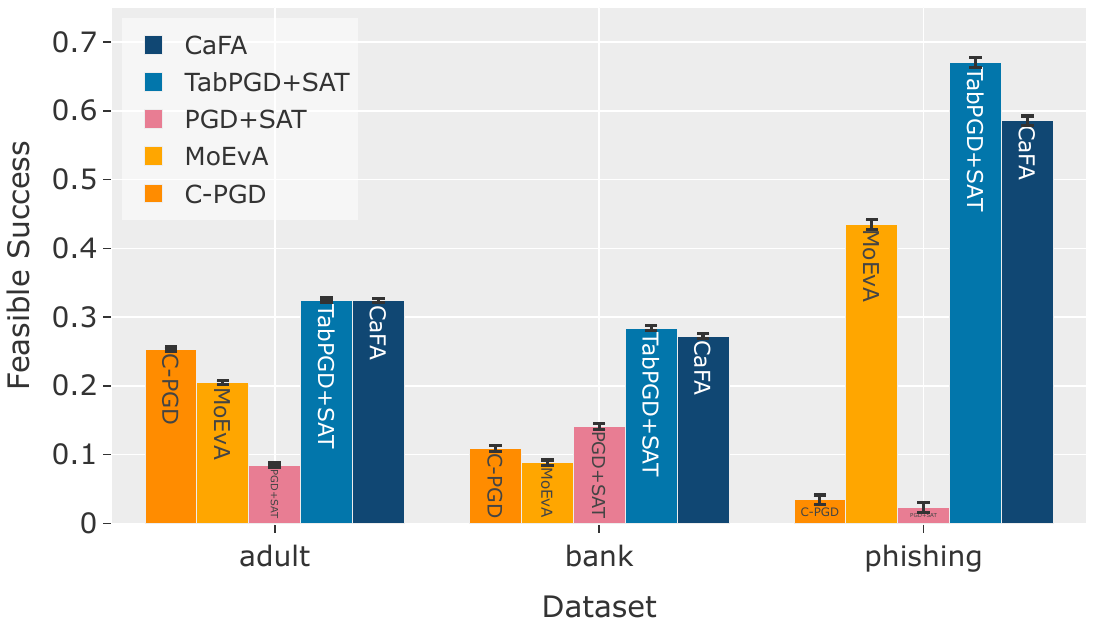}}
    \caption{\textbf{Full attacks} \label{figure:exp1-results-full}}
\end{subfigure}
\begin{subfigure}[t]{1.0\columnwidth}
    \centerline{\includegraphics[width=0.95\columnwidth]{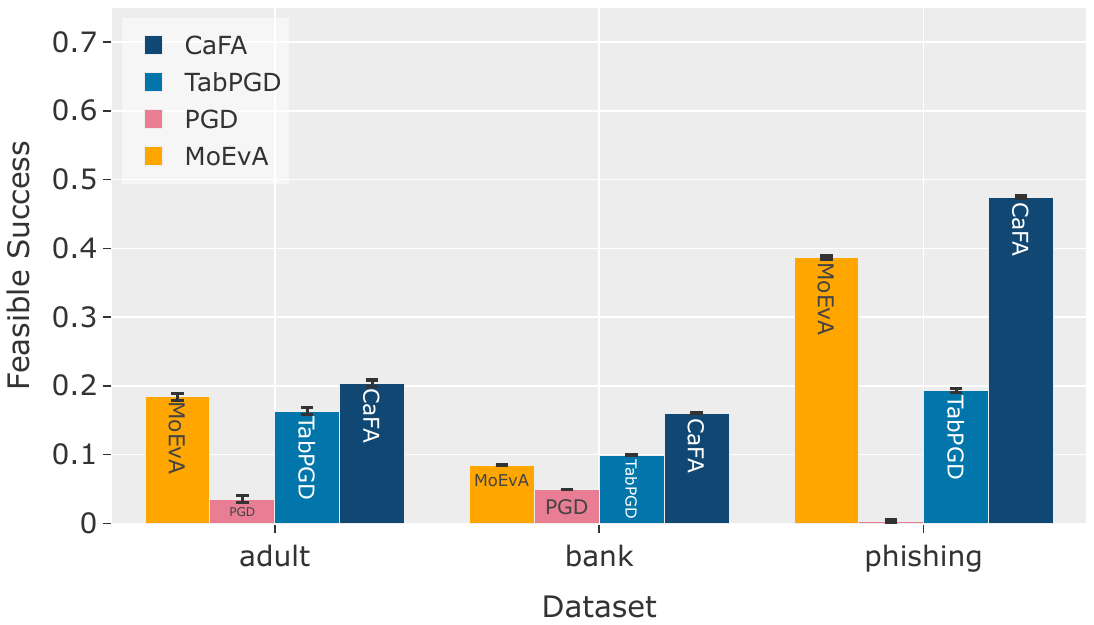}}
    \caption{\textbf{Ablation:} Disabling DCs access \label{figure:exp1-results-without-dcs}}
\end{subfigure}
    \caption{\label{figure:exp1-results}Comparison of the proportion
      of feasible and misclassified adversarial samples across
      attacks on MLPs.  
    }
\end{center}
\end{figure*}

We performed three experimental runs, each with a
different random seed and different 1K test samples.
We report the average results in \figref{figure:exp1-results} (more
details are included in \appref{app:exps:SuccDCs}).

As evaluations of the \textit{full} attacks show
(\figref{figure:exp1-results-full}),  
both variants of our \sys{} (\tabpgd{} and \tabpgd{}+\tabcw{})
attained higher feasible success rates than other attacks on all
datasets. We note a wider performance
gap (\figref{figure:exp1-results-full}, between \sys{} and other
attacks) compared to the gap shown when disabling DC access
(\figref{figure:exp1-results-without-dcs}), indicating \sys{} benefits
more from DC integrating. We also noticed varying performance across
datasets, which could be attributed to their distinct characteristics
(e.g., amount of categorical features). Nonetheless, \cafa{}
consistently found feasible adversarial samples in challenging
datasets like the \textit{bank} dataset, where other attacks showed
limited success.

As anticipated, attacks not integrating DCs were more likely to
produce out-of-domain adversarial samples, as indicated by the low
feasible success rates
(\figref{figure:exp1-results-without-dcs}). Still, across all
datasets, \sys{} consistently achieved the highest feasible success,
with MoEvA2 providing slightly a lower rate. This validates the
effectiveness of approaches incorporating structure constraints, and
more critically, providing perturbations with low $\ell_0$-norms. In particular,
these findings show that, in the absence of complex constraints,
minimizing perturbations' $\ell_0$-norms can improve the
likelihood of producing feasible adversarial samples. This is
presumably since altering fewer features reduces the interference with
feature dependencies. %

\subsubsection{Attacking TabNets} \label{subsection:exp1-tabnet}
To further validate our findings, we tested feasible success rates on
state-of-the-art transformer-based TabNet models.
A slight adjustment to \tabpgd{} was required to attack the
transformer-based models. Since TabNet uses a continuous embedding
layer to encode discrete categorical features, a special treatment was
required for their perturbation
(\secref{subsection:tech-tabpgd-attack}). Particularly, inspired by
the attack on binary malware by Kreuk et al.~\cite{kreuk2018},
we performed the following in each gradient-update step of the attack:
we calculated the continuous perturbation on the embedding
layer (as if these are continuous features we attack) using step size
maximizing attack success per line search, and picked the discrete
categories whose embeddings were closest (in Euclidean distance) to the
perturbed embedding.

\begin{figure}[t!]
\centering

\begin{subfigure}[t]{0.5\columnwidth}
\centerline{\includegraphics[width=1.0\columnwidth]{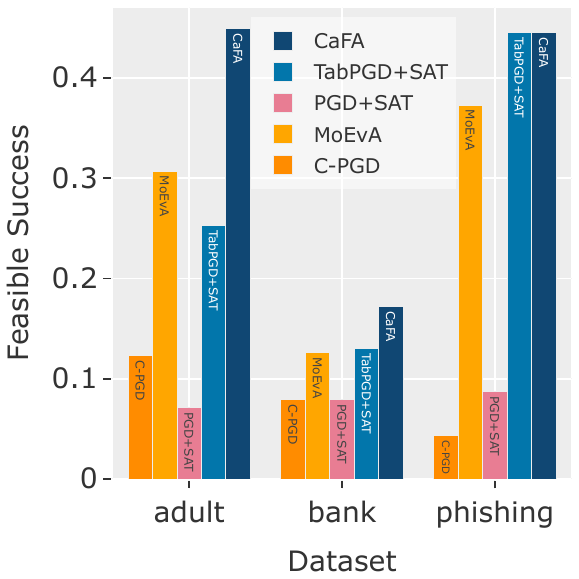}}
\caption{\label{figure:exp1-tabnet-soundness-phishing-with} Full attacks}
\end{subfigure}%
\begin{subfigure}[t]{0.5\columnwidth}
\centerline{\includegraphics[width=1.0\columnwidth]{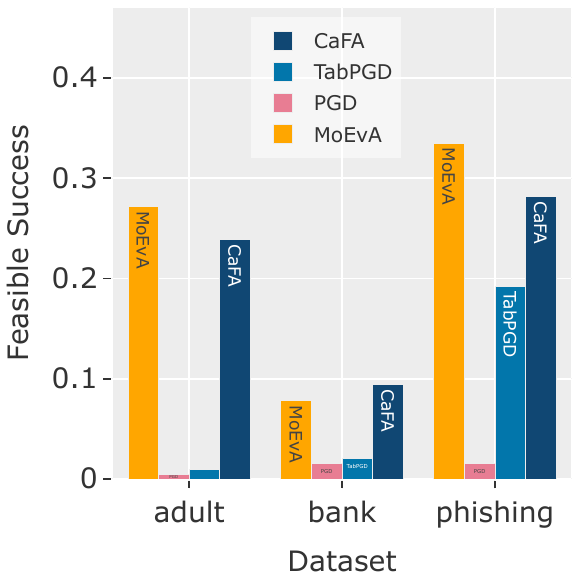}}
\caption{\label{figure:exp1-tabnet-soundness-phishing-without}
  Disabling DCs access}
\end{subfigure}

\caption{\label{figure:exp1-tabnet-soundness-phishing}
  \markdiff{Comparison of the proportion of feasible and misclassified
    adversarial samples across attacks on TabNets.}}
\end{figure}

From the results %
(\figref{figure:exp1-tabnet-soundness-phishing}), \markdiff{we observe that \sys{}'s feasible success surpasses prior attacks, with similar trends to those presented when attacking MLPs.} When disabling \dcs{} access, we notice a
decrease in the feasible success, with MoEvA2 outperforming \sys{} on
two datasets.
These results emphasize that \sys{} benefits from DC access more than other attacks do.

\subsection{Attack Cost} \label{subsection:exp-2-cost}

Next, we measured attacks' cost---a crucial aspect of attacks, besides
their feasible success rates
(\secref{section:threat-model}). We first inspect the costs of the different attacks (\textit{standardized-}$\ell_{\infty}$,$\ell_0$) while considering success, then we specifically examine \sys{}'s efficiency in reducing
perturbations' $\ell_0$-norms.

\begin{figure}[t!]
\begin{center}
\centerline{\includegraphics[width=0.9\columnwidth]{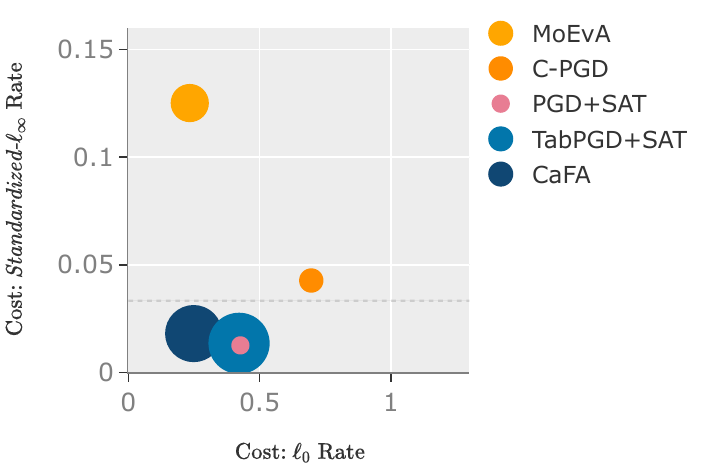}}
\caption{\label{figure:exp2-cost-2d}
  \markdiff{Comparing attacks' costs ($\ell_0$ and
    \textit{standardized}-$\ell_\infty$) while accounting for their
    feasible success (as indicated by the bubble size). 
  Dashed line marks the $\epsilon$ parameter used by \cafa{} and
  \tabpgd{}. Measures were averaged over all datasets (see
  \appref{app:addexps:cost} for per-dataset results). Larger
    bubbles closer to the bottom-left corner are better.}}
\end{center}
\end{figure}

\textbf{Comparing Attacks.} In this experiment, we ran attacks and
measured their costs and feasible success rates, while averaging these
over the datasets. We report the results in
\figref{figure:exp2-cost-2d}, and provide a more fine-grained
analysis of the \textit{standardized-}$\ell_\infty$- and $\ell_0$
costs in \appref{app:addexps:cost}. 

\markdiff{We find that our attacks provide the best trade-off between
  $\ell_0$- and \textit{standardized-}$\ell_\infty$-norms, while
  maintaining the highest feasible success. 
  This is opposed to other attacks,
  which either incur high costs in at least one norm  (MoEvA2, C-PGD),
  or fail to produce successful attacks (PGD).} 
We contend that using a standardized distance
metric is particularly useful for tabular data, where features’
value ranges may vary, thereby requiring different levels of
adversarial effort for an equivalent feature modifications.

\begin{figure}[t!]
\begin{center}
\centerline{\includegraphics[width=0.95\columnwidth]{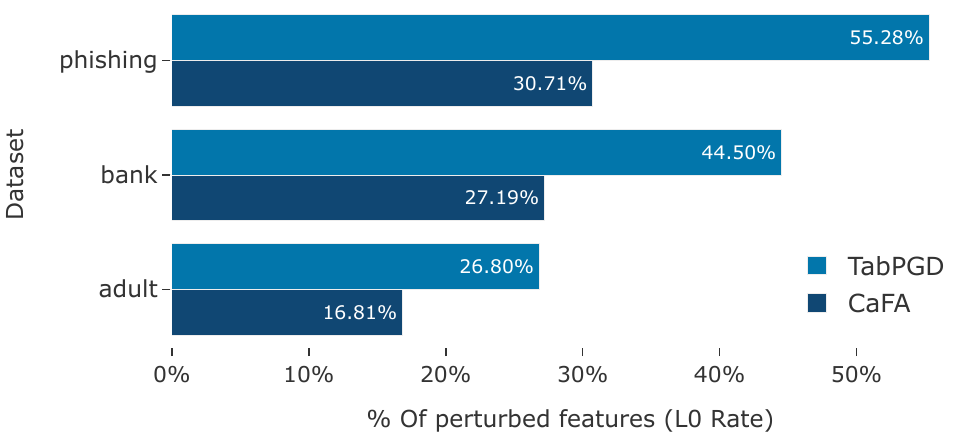}}
\caption{$\ell_0$ measures of the perturbations generated by
  feasible and successful
  adversarial samples of \sys{}'s variations (lower is better). To
  enable comparison across datasets, the measures were normalized by
  the number of features. 
  }
\label{figure:exp1-real-succcess-l0}
\end{center}
\end{figure}

\textbf{CaFA's $\ell_0$-norm.} We measured the perturbation's $\ell_0$-norm
of the feasible and successful adversarial samples crafted by \sys{}'s
variants. The results (\figref{figure:exp1-real-succcess-l0}) evidence
\tabcw{}'s importance---it can reduce up to half of the number of
perturbed features compared to \tabpgd{} alone. Overall, \sys{}
requires modifying 15--30\% of features, on average, for feasible
and successful attacks.

\subsection{Attack Run-Time} \label{subsection:exp-runtimes}

Run time is another factor that may impact \sys{}'s and other attacks'
adoption. To this end, we evaluated and compared attacks' average run
time for producing a single adversarial example. We executed attacks against the MLPs over 500 test
samples, one sample at a time, and calculated the average run time. We
then averaged the results across the three datasets.

\begin{figure}[t!]
\begin{center}
\centering
\centerline{\includegraphics[width=0.75\columnwidth]{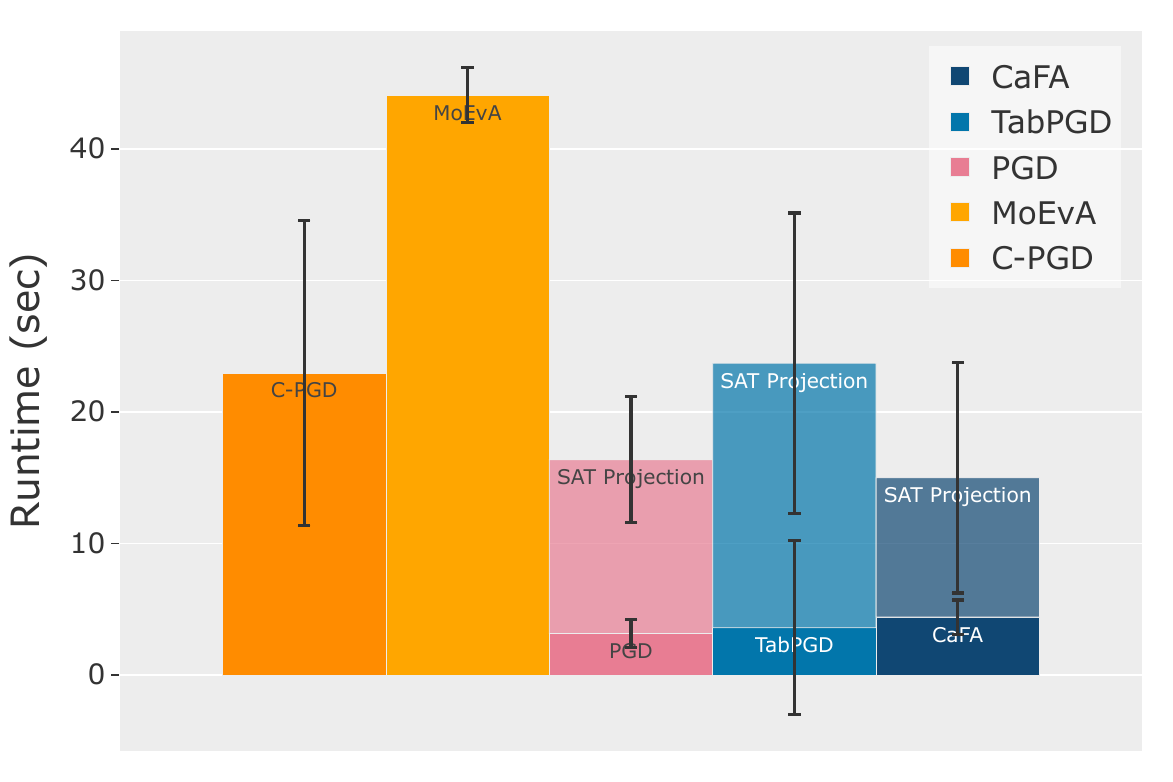}}
\caption{Average attack run-times per sample. For \sys{} (\tabpgd{}
  variants), we distinguish the perturbation stage from the projection.
}  
\label{figure:exp-runtimes} 
\end{center}
\end{figure}

As shown in \figref{figure:exp-runtimes}, both \sys{} variants are
roughly as fast as PGD, making them the fastest-measured 
methods incorporating constraints. A notable bottleneck in \sys{} is
the projection using SAT solvers. However, we observe
this bottleneck is alleviated when reducing the $\ell_0$ cost, as seen
with \tabcw{}. %

\subsection{Feasible Success With Valiant's Constraints} \label{exp-valiants}

Different constraint types can be integrated into \sys{}. To test its
flexibility, we explored incorporating Valiant's constraints, mining them as
described by Sheatsley et al.~\cite{sheatsley-constraints} (see
\secref{subsubsection:mining-valiants}). %
We compared the findings to
those attained by Sheatsley et al.'s~\cite{sheatsley-constraints}
attacks (PGD and CSP) on the \textit{phishing}
dataset. %
Due to the lack of publicly available implementation, we also tested
their PGD variants' performance with our attempted implementation.

\tabref{table:exp-sheatsley} presents the results
on the \textit{phishing} dataset (see \appref{appendix:valiants-study}
for results on other datasets). 
\sys{} yields $\sim$78\% successful and feasible adversarial samples
before projection, increasing to $\sim$88\% post-projection---about
$\times$1.5 more than the baseline attacks. Interestingly, Sheatsley
et al.\ previously concluded that Valiant's constraints markedly harms
attack success-rates~\cite{sheatsley-constraints}. Our findings draw a
different picture, with attacks succeeding relatively often even when
Valiant's constraints are accounted for. Results on the other datasets
further corroborate this finding
(\appref{appendix:valiants-study}). We attribute \sys{}'s improved
success rates mainly \markdiff{to how the attack handles heterogeneous features (e.g., step size) 
and to the minimization of the projected-features count, contributing to maintaining the samples misclassified.}

\begin{table}[t!]
    \centering
  \resizebox{1.0\columnwidth}{!}{%
  \begin{tabular}{l | rr | rr}
  \toprule

    & \multicolumn{2}{c|}{Full Attacks} & \multicolumn{2}{c}{Disabling Access} \\
    \textbf{Attack}  &   Comp.\ & Comp.\ \& Mis.\ & Comp.\  & Comp.\ \& Mis.\ \\ \midrule
    \sys{} (Valiant's)  &  100.0\% & \textbf{87.6\%} & 77.7\% & 77.7\%  \\ %
    PGD ~\cite{pgd-paper} (with tuning) & 100.0\% & 67.1\% & 3.3\% & 3.3\%  \\ %
    Sheatsley et al.'s PGD $(*)$~\cite{sheatsley-constraints} &  60.0\% & 60.0\% & 25.0\% & -  \\ %
    Sheatsley et al.'s PGD~\cite{sheatsley-constraints} &  89.4\% & 32.4\% & 2.8\%& 2.8\% \\ %
    Sheatsley et al.'s CSP $(*)$~\cite{sheatsley-constraints} &  100\% & 60.0\% & 80.0\% & -  \\
    \bottomrule
  \end{tabular}  
  }
   \caption{\label{table:exp-sheatsley} Attacking phishing detection
     with Valiant's constraints. We report the percentage of
     adversarial examples complying with the constraints and ones
     both complying and misclassified (i.e., feasible
     success rate) on the full attacks
     and when disabling Valiant's access.
    $(*)$ denotes results reported in prior
     work~\cite{sheatsley-constraints} (otherwise, the results were
     produced by our experiments).
  }
    
\end{table}

\subsection{Case Study: Evading Phishing Detection} \label{subsection:exp-case-study}

We ran a case study to challenge \sys{} with a real-world problem
space. Namely, we employed \sys{} to guide manual modification of
\emph{actual} phishing websites to evade detection by the
phishing-detection MLP
(\secref{subsecion:exp-setup-targeted-models}). Through this 
study, we critically analyzed different attack variants from a
problem-space perspective, highlighting the 
importance  of \sys{}'s attack objectives
(\secref{section:threat-model}) in a real-world scenario. 
\markdiff{We chose to focus on a set of \textit{phishing} samples, as
  this domain has clear and precise definition for
  \textit{implementability} (i.e., producing a valid HTML), in
  addition to up-to-date and realistic samples available
  online. %
} 

For the study,  we collected 11 samples of phishing websites \markdiff{in the wild} (i.e.,
HTML files), recently identified in the PhishTank archive \cite{phishtank-website} (sample IDs
found in \appref{appendix:practical-case-study}). To perform attacks, we first
extracted the website's features. %
Then, we employed \sys{} framework to generate an
adversarial feature-space instance. This instance subsequently served as a
\textit{recipe} for modifying the problem-space instances (original
URL and HTML code) to fool the targeted \ml{} detector.  
We repeated this process with different attacks, attempting to
manually implement the problem-space instances for each, and
annotating whether it was possible. 

\begin{figure}[t!]
\centering
\includegraphics[width=1.0\columnwidth]{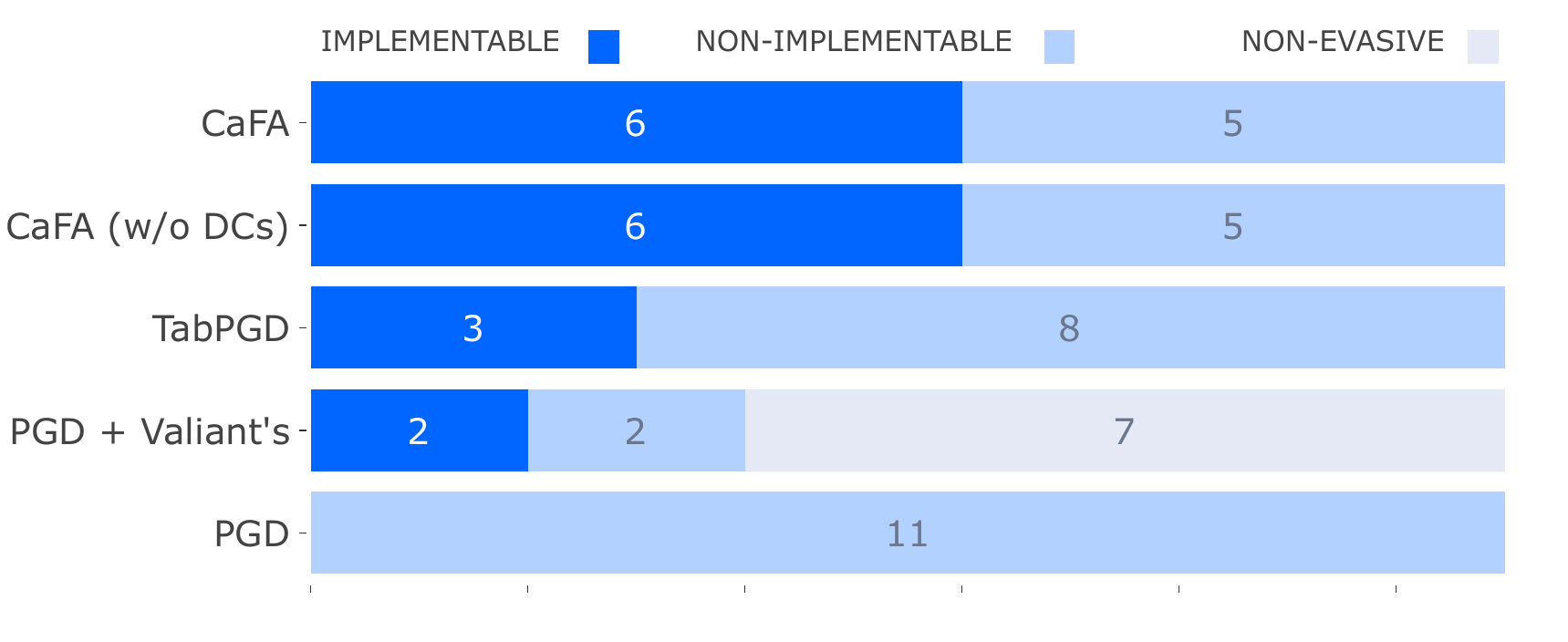}
\caption{\label{figure:case-study-stats} Results of manually examining
  11 samples of phishing websites originally detected by our
  phishing-detection MLP.
  We attempted to implement (the
  HTML and URL) the adversarial sample created by each attack (or at
  different stages of \sys{})
  and annotated whether it was implementable or not. 
  Some adversarial samples were also found to be non-evasive (i.e.,
  they were correctly classified). %
  }
\end{figure}

\markdiff{Using our attack, we successfully manipulated up to $\sim$55\% of the phishing websites to evade the model, while performing inconspicuous changes.} \figref{figure:case-study-stats} presents the result, and
\appref{appendix:practical-case-study} documents a problem-space
sample we produced.

\textbf{Attacks' Realizability.} Tailoring the attack to the
tabular domain markedly increased attacks' success (PGD vs.\ \sys{} in
\figref{figure:case-study-stats}) and led to better alignment with
problem-space constraints compared to prior work (PGD+Valiant's
vs.\ \sys{} in \figref{figure:case-study-stats}). 
Enforcing structure
constraints (\tabpgd) has immediately made some
attacks practical by eliminating critical feature inconsistencies,
such as ensuring an \textit{integer} URL in \textit{reasonable}
length. Additionally, minimizing the number of perturbed features 
(i.e., $\ell_0$-norm minimized by \tabpgd{}+\tabcw{} even without
incorporating DCs in \sys{}), originally motivated by cost, also
resulted in implementable adversarial samples in the feature
space. This result is consistent with those show in
\secref{subsection:exp1-real-success}, where minimizing $\ell_0$-norm
already improved feasible success rates, even without integrating
DCs.

\textbf{Direction for Improving Realizability.}
The results show that, in the particular use-case examined,
integrating the DCs mined with FastADC~\cite{fastadc} did not provide
an improvement in realizability compared to the attack variant
merely minimizing the perturbations' \lpnorm{0}-norms (\cafa{} vs.\
\cafa{} (w/o \dcs{}) in \figref{figure:case-study-stats}). 
A closer inspection showed that the mined set of \dcs{} lacked highly
nuanced constraints---for instance, subtle dependencies between
different proportions of hyperlinks, akin to the last golden
constraint presented in \tabref{table:phishing-golden-constraints} (\appref{exp0-sound-vs-comp-phishing}), %
were not mined---as opposed to the coarser
dependencies the \dcs{} aptly identified
          (e.g., numerical character count in a URL  not exceeding its
length). This is not surprising, due to imperfect completeness and
soundness measures of DCs (\appref{exp0-sound-vs-comp-phishing}). 
Accordingly, projecting on the mined DCs still led to violations of
constraints imposed by the problem space, thus harming
realizability. %
While \dcs{} posses of high expressivity, capable of capturing
nuanced dependencies, a profound challenge lays in mining them to
capture genuine constraints with high completeness and soundness.
Thus, enhancements in mining \dcs{}, leading to higher quality
constraints in the future, would also boost \sys{}'s efficacy.

\textbf{Cost Efficiency.}  Throughout the manual translation of perturbations
into problem-space modifications, the significance of cost was
apparent. 
The bounded \textit{standardized-}$\ell_\infty$-norm in \sys{}
 has contributed to limiting the effort required for
 certain modifications. For example, it meant that only minor character
 alterations were required for the URL. A drastic URL change would
 necessitate significant effort, such as potentially purchasing a new
 domain. Additionally, minimizing $\ell_0$-norm helped in narrowing down the
 \textit{variety} of efforts required from the attacker. For instance,
 modifying only \textit{href} HTML tags was simpler than tampering with
 multiple, potentially interdependent, HTML logics. 
 \markdiff{In this study we also observed that minimizing the
   adversarial effort has served the implementable adversarial samples
   in maintaining their functionality (i.e., phishing pages kept
   identical appearance when implemented after evasion).}

%% file: tables/bank-golden-constraints.tex
\begin{table} [t!]

  \resizebox{\columnwidth}{!}{%
  \begin{tabular}{l r}
  \toprule
    \bf Constraint & \bf Comp.\ Rate (\%) \\ 
    \midrule
    $x.previous = 0 \implies x.poutcome = -1$ & 100.0\% \\  
    $x.previous = 0 \impliedby x.poutcome = -1$ & 100.0\% \\ 
    $x.job = \text{`student'} \implies x.marital = \text{`single'} \land x.age \leq 35 $ & 99.8\% \\ 
    $x.job = \text{`admin'} \implies x.education = \text{`secondary'} $ & 100.0\% \\ \bottomrule
  \end{tabular}
  }
    \caption{\label{table:bank-golden-constraints}Golden constraints
      derived by Deutch and Frost~\cite{daniel-constraints} from the
      \textit{bank} dataset. We describe each constraint relative to a feature
      vector $x\in X$ and report the precentage of samples complying with
      it. %
    }
\end{table}

%% file: tables/exp0-comp-vs-sound-picks.tex
 \begin{table}[t!]
 \centering
  \resizebox{1.0\columnwidth}{!}{%
  \begin{tabular}{l | r r r|r r r|l} 
    \toprule
    \bf Type & \bf \textit{$n_{dcs}$} &\bf  \textit{$n_{tuples}$} & \bf \# Constr. & \bf Compl. & \bf Sound. & \bf F1 & \bf  Notes \\ \midrule
    
    Valiant's  & - &  - & 51K & \textbf{99.1\%} & 7.9\% & 14.7\%  &  \\ \midrule
    
    DCs  & 5000 & 1 & 5K & 88.3\% & 75.1\% & \textbf{81.2\%} & Max.\ F1, DCs \#1 \\ %
    
    DCs  & 7000 &  500 & 3.5M & 14.4\% & \textbf{100.0\%} & 25.2\% & DCs \#2 \\  %
    
    DCs  & 100 &  1 & 0.1K & \textbf{99.1\%} & 7.1\% & 13.3\% & DCs \#3 \\ %
    
    \bottomrule
  \end{tabular}}
    \caption{\label{table:com-sound-picks} Constraint sets of interest and their empirical soundness, completeness, and F1 scores. \dcs{} differ in their configuration (i.e., \textit{$n_\mathit{dcs}$} and \textit{$n_\mathit{tuples}$}). %
    }
\end{table}

%% file: discussion.tex
\section{Discussion} \label{section:defenses}

\markdiff{We now discuss potential defenses against \sys{} and other
  future directions.} 

\textbf{Defenses.} \markdiff{Adversarial training, demonstrated
  effective under various threat models}
(e.g.,~\cite{android-bostani-2023,simonetto-2022,
  pgd-paper}). \markdiff{It 
  is perhaps the most natural defense against adversarial examples
  generated by \sys{}. To conduct adversarial training, one would need
  to execute \sys{} for each training batch to craft adversarial
  examples for training. Doing so would incur a significant
  training-time overhead} (\figref{figure:exp-runtimes})
\markdiff{potentially rendering adversarial training infeasible. Thus,
  speeding up attacks (e.g., by accelerating, or relaxing, the
  projection) could be critical for enabling adversarial training.} 

\markdiff{Yet another interesting direction to explore would be detecting
  adversarial examples crafted by \sys{}. In particular, due to the
  nature of projections with SAT solvers, we conjecture that, compared
  to benign samples, \sys{}'s adversarial examples would violate
  constraints by changing more features by smaller
  amounts. Accordingly, if such differences exist, statistical
  techniques would allow telling adversarial and benign samples
  apart. These techniques could supplement pre-existing detectors}
(e.g., \cite{roth-2019-detect})\markdiff{, enhancing their detection
  accuracy.}  

\markdiff{Extending a concept proposed by Simonetto et
  al.}~\cite{simonetto-2022}, \markdiff{another approach is enriching
  the model's input with more features computed as complicated
  functions of other features, with the goal of inducing a challenge
  for constraint mining. For example, one may add a feature via a
  complicated arithmetic function of other features that cannot be
  represented by DCs, despite their expressivity.} 

\markdiff{Finally, as ML models interoperate with other modules (e.g.,
  expert-defined rule-based tools for phishing detection) when
  deployed in real-world systems, the complexities of such
  systems may pose natural challenges to adversaries. Therefore, an
  intriguing direction for future research
  would be to evaluate the robustness of complex real-world systems
  against \sys{} and similar attacks.}

\textbf{Future Work.} %
Besides enhancing models' robustness against \sys{} and related
attacks, it would be interesting to explore avenues to improve
\sys{}'s performance and further generalize it. For instance, 
improved projection techniques can help improve \sys{}'s
performance, possibly by relaxing the requirement to satisfy \textit{all}
the constraints to \textit{most} of them, or explicitly accounting 
for cost and misclassification objectives. Additionally, 
to make it more widely applicable, it would also be helpful to extend 
\sys{} to non-\nn{}-based \ml{} models.

%% file: conclusion.tex
\section{Conclusion} \label{section:conclusion}

We presented \sys{}, a system producing adversarial examples to
mislead \ml{} models classifying tabular data. \sys{} tackles the two
primary challenges of realizability---i.e., adversarial perturbations
in the feature space may not be implementable in the problem
space---and cost minimization---i.e., ensuring adversaries' effort is
minimized when crafting adversarial artifacts. To tackle the former, \sys{}
leverages structure constraints and automatically mined denial
constraints, ensuring that adversarial perturbations comply with
domain-imposed restrictions. For the latter, \sys{} seeks to minimize
the number of features perturbed as well as the extent to which they
are altered while accounting for the heterogeneity of features. The
empirical evaluation with three commonly used datasets and two
standard models demonstrate, among others, the
\textit{(1)} advantages of the constraints used by
\sys{} (specifically, better balancing soundness and completeness than
previously used constraints);
\textit{(2)} \sys{} superiority at generating feasible adversarial
examples that are misclassified while satisfying integrity
constraints compared to prior attacks; and
\textit{(3)} \sys{} ability to minimize the number of feature's
perturbed and the perturbations' magnitude. We open-source
\sys{}’s implementation, hopefully it would be used as a
generic mean for evaluating tabular classifiers’ robustness
against practical attacks prior to deployment.

%% file: acknowledgements.tex
\section*{Acknowledgements}

This work
has been partially funded by the European Research
Council (ERC) under the European Union's Horizon 2020 research and
innovation programme (grant agreement No.\ 804302)
and has been supported in part
by a grant from the Blavatnik Interdisciplinary Cyber Research Center
(ICRC);
by Intel\textregistered{} via a Rising Star Faculty Award;
by a gift from KDDI Research;
by Len Blavatnik and the Blavatnik Family foundation;
by a Maof prize for outstanding young scientists;
by the Ministry of Innovation, Science \& Technology, Israel (grant
number 0603870071); 
by a gift from the Neubauer Family foundation;
by NVIDIA via a hardware grant;
and by a grant from the Tel Aviv University Center for AI and Data
Science (TAD).

%% file: appendix.tex
\label{section:appendix}

\section{Additional Technical Details}

\subsection{Valiant's Constraints} \label{subsubsection:mining-valiants}

Inspired by Sheatsley et al. \cite{sheatsley-constraints}, we consider
the boolean constraints mined by Valiant's PAC learning
algorithm~\cite{valiants-paper-1984} to capture relation constraints.
Specifically, as described next, we focus on a particular variant of
Valiant's constraints.

\textbf{Definition.}
Valiant's algorithm learns boolean constraint theories from the
data~\cite{valiants-paper-1984}. It captures constraints 
as Conjunctive Normal Form (CNF) logical
formulas, %
where each predicate enforces the equality of some feature to a
constant. %
Following Sheatsley et al., we use Valiant's algorithm to mine
permissive constraint theories,\footnote{Specifically, we set $k$=1}
enabling us to evaluate robustness against the least constrained
adversary. Doing so also renders the mining process tractable, as
mining less permissive constraint theories with Valiant's 
algorithm is often intractable in practice.

\textbf{Binning.} The complexity of the Valiant's algorithm is
exponential in the size of the features' support (i.e., values they
can admit). In practice, many features have large support
(e.g. continuous features may admit as many distinct values as the
number of samples). Hence, to make Valiant's algorithm feasible to
run, we discretize continuous features, similarly to prior
work~\cite{sheatsley-constraints}. Particularly, we discretize
continuous features by binning their values into one
of \(k_\mathit{bin}\) bins formed by applying the $k$-means
algorithm\footnote{We
use \texttt{sklearn}'s \texttt{KBinsDiscretizer} with k-means.}
to each feature's support.
We note that $k$-means led to the best observed performance, higher
than the OPTICS algorithm considered in prior
work~\cite{sheatsley-constraints}. 
We denote the discretized support of feature $i$
as \(\widetilde{S}_i\) (a discrete set of
size \(k_\mathit{bin}\)).\footnote{Amount of bins per dataset:
$k_\mathit{bin}$=4 for \textit{bank}, $k_\mathit{bin}$=6 for \textit{phishing},
and $k_\mathit{bin}$=4 for \textit{adult}.}  For
features that were not discretized, \(\widetilde{S}_i\) simply stands
for the feature's actual support. We denote the discretized
feature space \(\widetilde{X}\). 

\textbf{Parameterization.} %
Our implementation runs over the following form of possible constraints: 
    \[
    \Gamma := 
    \left\{ \forall x\in X.\lor_{i\in{[d]}}\left(x_{i}=s_{i}\right)\mid s_{1}\in \widetilde{S}_{1},\dots,s_{d}\in \widetilde{S}_{d}\right\} 
    \]
Put simply, each constraint, defined by a support
vector, \((s_1, \dots s_d)\), requires that each sample \(x\) in
the feature space would have at least a single
coordinate, \(j\in[d]\) , where it identifies with the values of
the support vector (i.e., \(x_j = s_j\)). 
As noted to earlier, 
other parameterizations of Valiant's constraint space consider
a \textit{set} of possible values per clause
(e.g., a clause can be of form $x_i \in S'_i$ instead of $x_i = s'_i$),
leading to an exponential growth in the mining complexity.

\textbf{Mining Process.} %
Given the training set  after
discretization, \(\widetilde{X}_{train}\), Valiant's algorithm begins
with \(T:=\Gamma\), the space of all possible constraints, and returns
returns \(T\), the set of the mined constraints~\cite{valiants-paper-1984}. 
The algorithm iterates on each (discretized)
sample $\widetilde{x} \in \widetilde{X}_{train} $, and checks,
for each potential constraint \(t\in T\), whether \(\widetilde{x}\)
satisfies it or not. If \(\widetilde{x}\) does not satisfy \(t\) ,
then \(t\) is discarded from \(T\). 
In the worst case, when no constraint is discarded, we get a time
complexity of
\(O\left(\left|\widetilde{X}_{train}\right|\Pi_{i=1}^{d}\left|\widetilde{S}_{i}\right|\right)\).

\textbf{Advantages of \dcs{} Over Valiant's.} Besides the empirical
evidence above, we opt for \dcs{} for their desirable properties over
Valiant's constraints. First, \dcs{} using soft constraints
can potentially identify valuable constraints that may be
excluded by Valiant's due to a few noisy
sample. Furthermore, as opposed to Valiant's, mining \dcs{} does not
necessitate data discretization and is applied to the raw form of the
samples. This absence of
discretization enhances the representativeness of the constraints; for
example, Valiant's fails to model dependencies within the discretized
ranges, which could be vital for addressing small
perturbations. Finally, the ability to work directly with raw data
also enables \dcs{} to incorporate additional, important assertions on
the literals, such as setting precise perturbation bounds.

\subsection{Ranking \dcs{}} \label{appendix-dcs-ranking}

As mentioned in \secref{subsection:mining}, in our framework, we use
only a subset of the mined \dcs{}, based on a heuristic ranking
scheme. %
Here, we elaborate on this scheme, establishing it based on metrics
from data-integrity literature. 

In this work, we use the following standard metrics %
that quantify different aspects of \dcs{} (e.g., the
``interestingness'' constraints~\cite{discovering-dc} as well as their
satisfaction by in-domain samples~\cite{approximate-dcs-2019,
adcminer-paper, fastadc}):
\begin{itemize}
    \item  \textbf{Succinctness}~\cite{discovering-dc}.  Following
    ``Occam's Razor,''  an insightful DC would be short and
    concise. This metric stands for how close the DC's amount of
    predicates to the minimum length of predicates. The closer, the
    better. 
    \item  \textbf{Coverage}~\cite{discovering-dc}. %
    Even if a DC is satisfied, the amount of
    satisfied inner-predicates varies. This amount of satisfied
    predicates can be seen as the support in the data. %
    \textit{Coverage} measures, for a single DC, the
    weighted average of the amount of satisfied predicates, over the
    different sample pairs in the data. 
    \item  \textbf{Pairs violation}~\cite{approximate-dcs-2019,
    adcminer-paper, fastadc} measures the proportion of sample pairs violated by
    the DC (out of all sample pairs). Also known as
    the \textit{violation rate} when considered
    in \textit{soft} \dcs{}. 
    \item  \textbf{Sample violation}~\cite{adcminer-paper} quantifies
    the proportion
    of samples for which exist \textit{any} pair of that violates the
    DC, over all samples. 
\end{itemize}

To use these metrics in our ranking scheme, the following process is applied after mining the \dcs{}:
\begin{enumerate}
    \item We sample a subset (3K) samples from the training set.
    \item We select (10K) the \dcs{} to rank, according to an
    efficiently calculated metric--succinctness.
    \item Each DC is assigned a score according to a linear
    combination of the four metrics. The coefficients for the linear
    combination are selected by manually inspecting the ranking
    provided by different values.
    \item For each DC, we choose the most satisfying \textit{other}
    tuples, calculated as the satisfaction rate of
    the \dcs{} with the %
    \textit{other} tuple. 
    \item We choose the $n_\mathit{dcs}$ top-ranked \dcs{}, and, for each, we
    choose the $n_\mathit{tuples}$ top-ranked \textit{other} tuples. 
\end{enumerate}

\section{Additional Experimental Setup Details}

\textbf{MoEvA2's Norm Function} \label{appendix:moeva-norm-choice}
Throughout our experiments (\secref{section:experiments}), when using
MoEvA2~\cite{simonetto-2022}, we set it to run while minimizing
the \lpnorm{2}-norm, although we are measuring a variation of
\lpnorm{\infty}-norm. This choice was taken after finding %
that the feasible success of MoEvA2 is higher under this choice of
norm function.

\begin{figure}[t!]
\begin{center}
\centerline{\includegraphics[width=0.9\columnwidth]{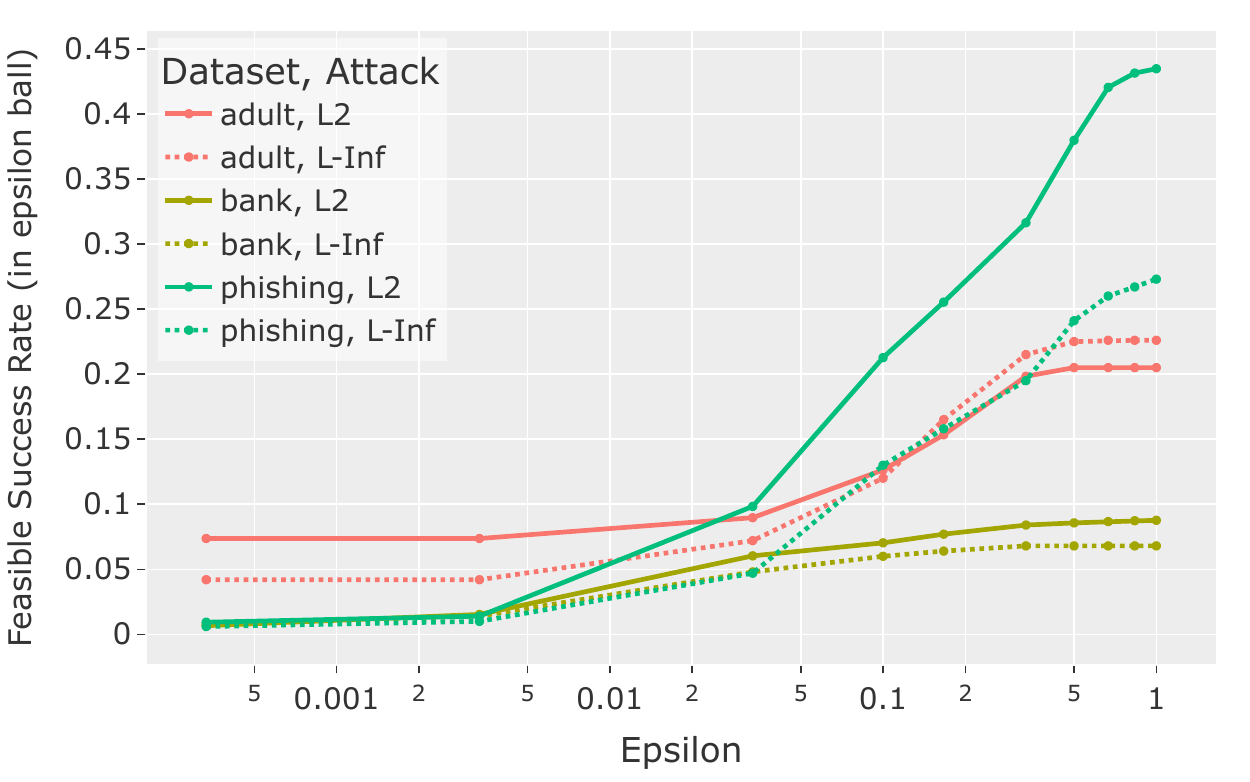}}
\caption{Comparing the feasible success rates of MoEvA2 with \dcs{}
under \lpnorm{\infty}- and \lpnorm{2}-norms.}
\label{figure:moeva-norm-choice}
\end{center}
\end{figure}

We ran an experiment similar to our original analysis (\secref{subsection:exp-2-cost}), and compare the performance of two instances of MoEvA2, one with the distance objective of $\ell_2$ distance and the other one with the $\ell_\infty$ objective. We report the result in \figref{figure:moeva-norm-choice}, where we observe that, although not intuitive, incorporating the $\ell_2$ objective provides better performance under our standardized-$\ell_\infty$ metric.

\newpage
\section{Additional Experimental Results}

\subsection{Cost Analysis}
\label{app:addexps:cost}

\begin{figure}[t!]
\begin{center}
\centerline{\includegraphics[width=0.9\columnwidth]{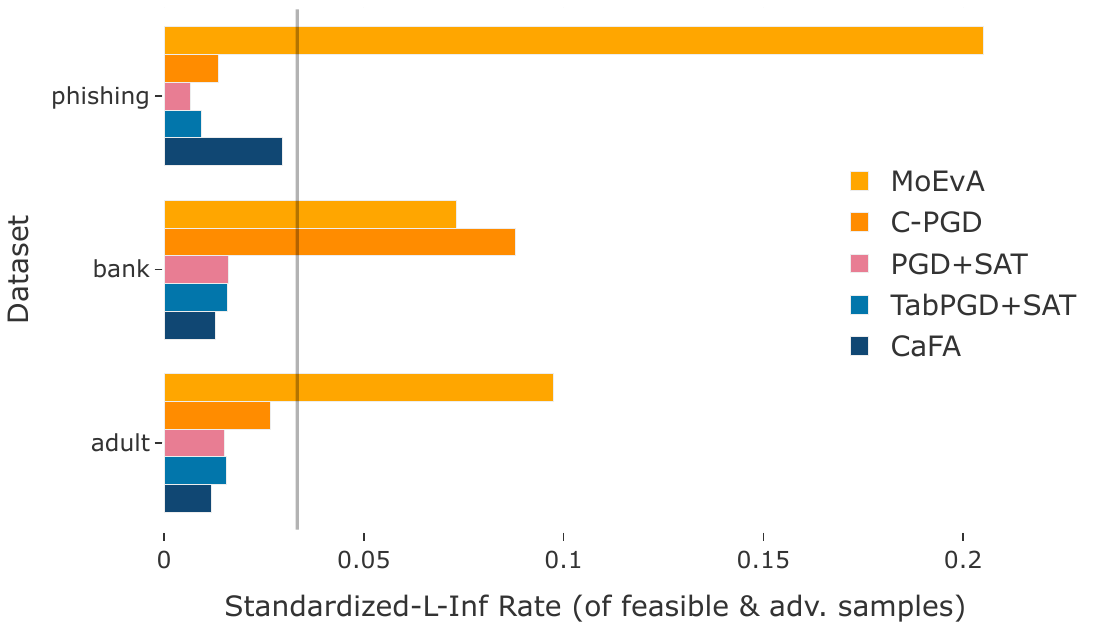}}
\caption{
\markdiff{The \textit{standardized-}$\ell_{\infty}$ cost of
feasible and successful adversarial samples %
(lower is better). The vertical line corresponds to CaFA's cost bound
($\epsilon=\frac{1}{30}$).}
}
\label{figure:real-succcess-l-inf-bars}
\end{center}
\end{figure}

\begin{figure}[t!]
\begin{center}
\centerline{\includegraphics[width=0.9\columnwidth]{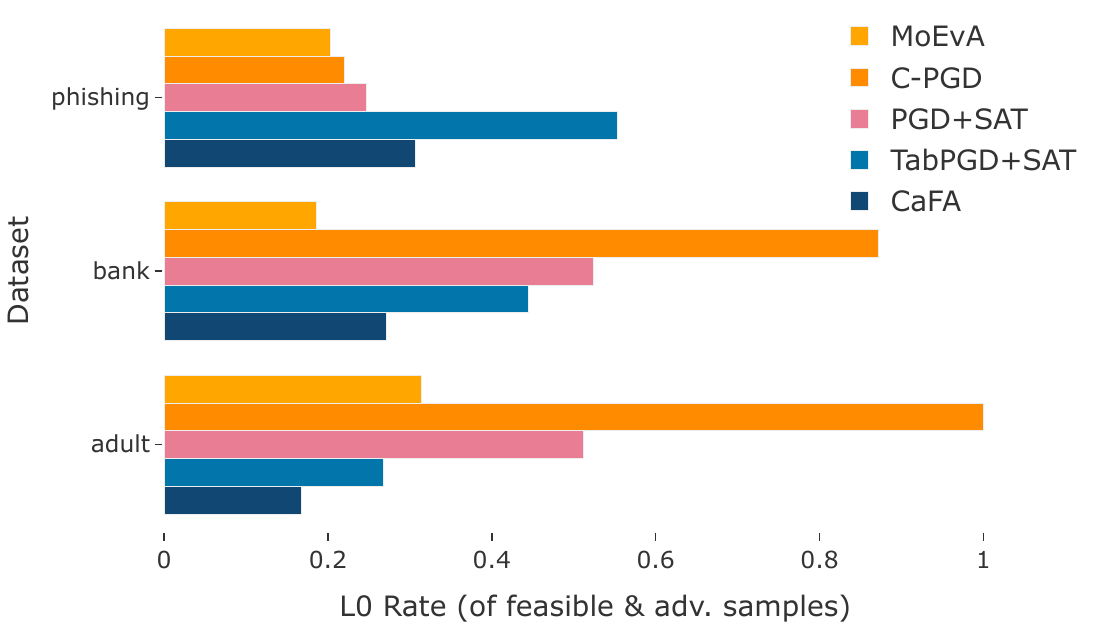}}
\caption{
\markdiff{The $\ell_0$ cost of feasible and successful adversarial
samples (lower is better). %
To enable cross-dataset comparisons,
we normalized values by the number of features.}
}
\label{figure:real-succcess-l0-bars}
\end{center}
\end{figure}

Here, we include a more fine-grained comparison of attack costs. The
measures were taken on the subset of successful and feasible
adversarial samples, having considered the trade-off between the
feasible success rate and the costs
in \secref{subsection:exp-2-cost}. We report the comparison over
the \textit{standardized-}$\ell_\infty$ cost
in \figref{figure:real-succcess-l-inf-bars} and over the $\ell_0$ cost
in \figref{figure:real-succcess-l0-bars}.  

\subsection{Impact of Constraint Choice}
\label{subsection:exp-sound-vs-success}

\begin{figure}[t!]
\centering
\centerline{\includegraphics[width=0.85\columnwidth]{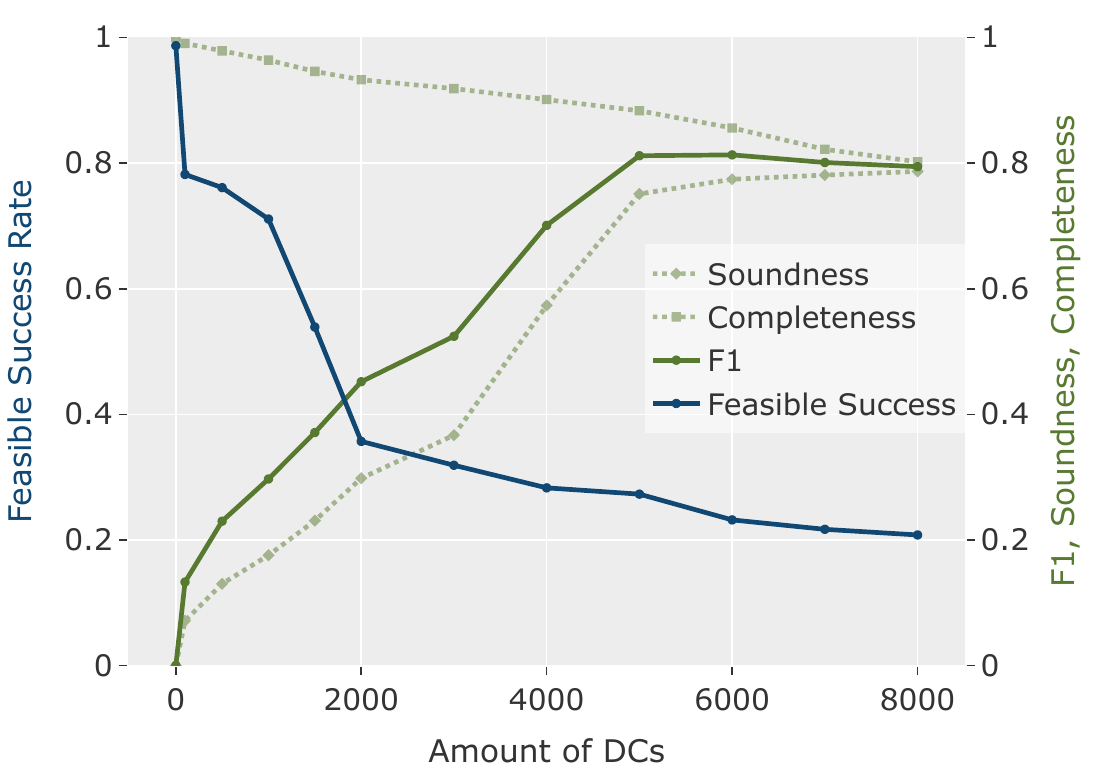}}
\caption{Constraints' quality and \sys{}'s feasible
  success rates for different choices of DC sets on the \textit{bank}
  dataset. %
}
\label{figure:exp3-soundness}
\end{figure}

\textbf{Bank Dataset.}
A key trade-off lies between the feature-space constraints  and
attacks' feasible success. In general, the larger the constraint sets,
the more challenging attacks are. To better understand this
relationship, we explored how attacks' feasible success rates
behave when varying the number of constraints (thus, also attaining
different soundness-completeness tradeoffs). We first ran the
analysis on the \textit{bank} dataset,
running \sys{} with varying constraint sets by controlling the
$n_\mathit{dcs}$ parameter. For each constraint set, we measured
feasible success, empirical soundness, and completeness. The results are shown
in \figref{figure:exp3-soundness}.

We observed that, \textit{within} the considered set sizes, the
constraint set quality (indicated by the F1 score) mostly increases
with the set size, reaching its highest value when considering 5K
constraints, the number of constraints used in most of our
experiments.
This increase in F1 scores correlates with a decrease in attack performance. For
instance, when using a constraint set with a 13\% F1 score, \sys{}
achieved a feasible success rate of  $\sim$78\%.
Conversely, a higher quality constraint set with an 80\% F1 score
leads to a
decline of feasible success rate to $\sim$30\%. These findings emphasize the
challenging landscape posed by high-quality constraints.

\textbf{Phishing Dataset.}
\label{exp0-sound-vs-comp-phishing}
Thus far, we have evaluated how the choice of constraints impact
soundness, completeness, and feasible success rates on
the \textit{bank} dataset.
To further validate the findings, we report on a similar evaluation
with the \textit{phishing} dataset.
In this experiment, we used the exact setting and measures considered
for the \textit{bank} dataset. However, as no golden constraints were
derived for the \textit{phishing} dataset in prior work, we derived
those manually (see  \tabref{table:phishing-golden-constraints}).

\begin{figure}[t!]
\centering
\centerline{\includegraphics[width=0.85\columnwidth]{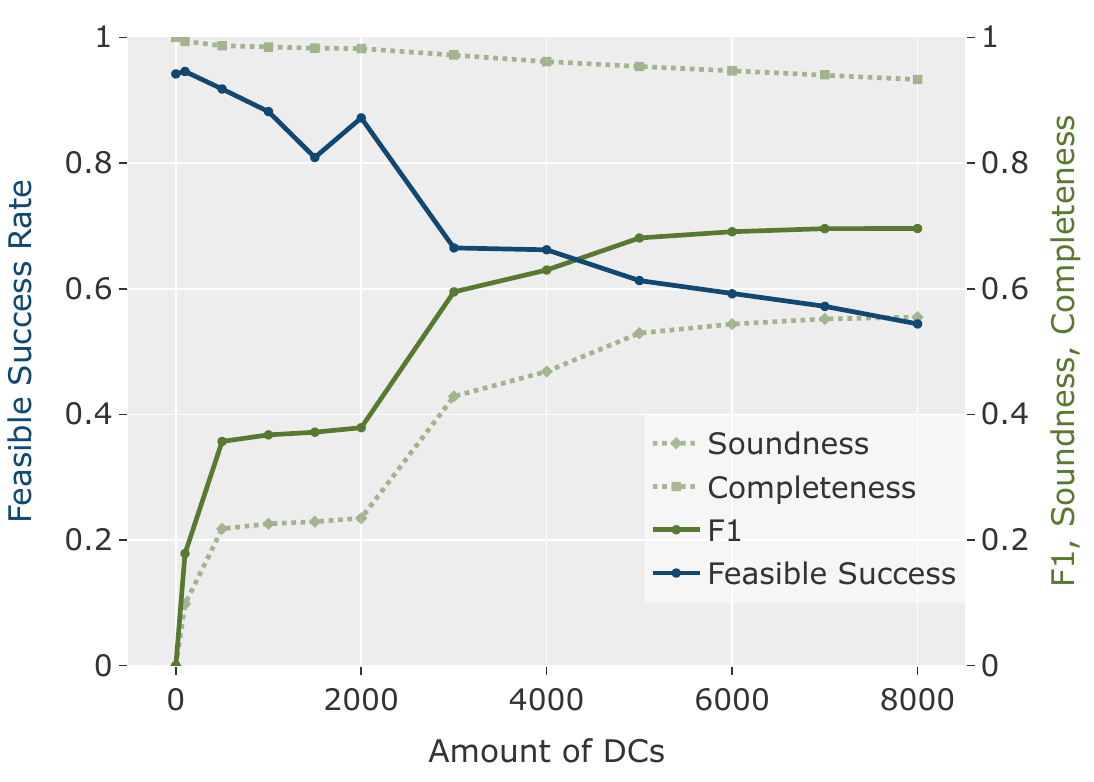}}
\caption{Constraints' quality and \sys{}'s feasible
  success rates for different choices of DC sets on the \textit{phishing}
  dataset.}
\label{figure:exp3-soundness-phishing}
\end{figure}

\figref{figure:exp3-soundness-phishing} reports the results.
Consistent with the trends observed with \textit{bank} dataset, we
note that the F1 score increases as the size of the constraint set
expands. We also observe an inverse relationship between the size of
the constraint set and the attack's feasible success. Specifically,
this evaluation further justifies our choice of \dcs{} configuration,
as determined by our analysis of the \textit{bank} \dcs{}
(\secref{subsection:exp0-sound-vs-comp})---this configuration, with
$n_\mathit{dcs}$=5K and $n_\mathit{tuples}$=1 led to the (roughly)
highest of F1 score on the \textit{phishing} dataset too.

\begin{table} [t!]
\centering
  \resizebox{1.0\columnwidth}{!}{%

  \begin{tabular}{r l r}
  \toprule
    \textbf{ID} & {\bf Constraint} & \textbf{Comp. Rate (\%)} \\  \midrule
    \#1 &$x.NumNumericChars < x.UrlLength$ & 100.0\% \\ %
    \#2 &$(x.NumSensitiveWords * 2)  < x.UrlLength$ & 100.0\% \\ %
    \#3 &$x.PctNullSelfRedirectHyperlinks = 1 \implies x.PctExtHyperlinks = 0$ & 100.0\% \\ %
    \#4 &$(x.PctNullSelfRedirectHyperlinks + x.PctExtHyperlinks) \leq 1$ & 99.9\% \\ %
    \multirow{2}{*}{\#5} &$x.PctExtNullSelfRedirectHyperlinksRT = 1 \implies $ & \multirow{2}{*}{100.0\%} \\ %
           & $x.PctNullSelfRedirectHyperlinks + x.PctExtHyperlinks < 0.3$
    &  \\ \bottomrule
  \end{tabular}}
    \caption{\label{table:phishing-golden-constraints} Manually
  derived golden constraints for the \textit{phishing} dataset. We
  describe each constraint relative to a feature-space vector $x\in X$,
  and report the percentage of samples from the dataset that
  comply with it.}
\end{table}

\subsection{Case Study: Evading \textit{Phishing} Detector} \label{appendix:practical-case-study}

\begin{figure}[t!]
\centering
\centerline{\includegraphics[width=0.7\columnwidth]{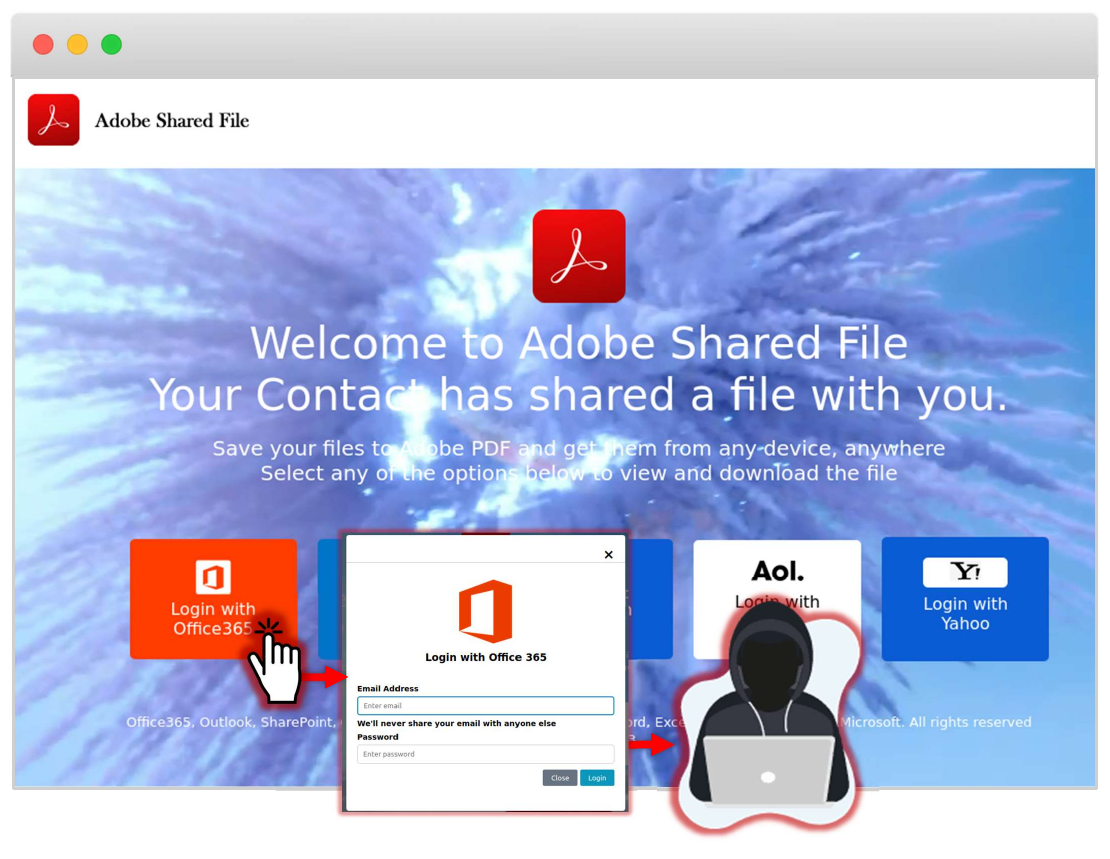}}
\caption{A phishing page for which we demonstrate an evasion. The page
was taken PhishTank (ID: 8302119). The page tricks users to login to
various services, such as Office 365, and sends their credentials to
a malicious server.}
\label{figure:phishing-website-example}
\end{figure}

In this study, exemplifying one of 11 samples\footnote{PhishTank sample IDs:
  8302119, 8307185, 8314314, 8309078, 8309085, 8310709, 8309529,
  8309492, 8313454, 8313452, 8314312.} inspected in the case study
  (\secref{subsection:exp-case-study}), we demonstrate a translation
  of adversarial example attacks from the feature-space to the
  real-world problem-space, using a real phishing website as an
  example. Due to the absence of raw data of the \textit{phishing}
  dataset \cite{phishing-dataset}, we rely on a recently identified
  phishing website taken from the PhishTank archive
  (\figref{figure:phishing-website-example}). Using this sample, we
  highlight the importance of \sys{}'s threat model
  (\secref{section:threat-model}) in a realistic scenario. 

First, we extracted the website's features, by implementing the
feature-extraction logic (the dataset does not include code for
feature extraction). Subsequently, we employed \sys{} to generate an
adversarial sample in the feature space
(see \tabref{table:example-case-study}). This sample served as
a \textit{recipe} for modifying the original URL and HTML code to fool
the \ml{}-based detector.

 \begin{table}[t!]
 \centering
  \resizebox{1.0\columnwidth}{!}{%
  \begin{tabular}{l|rrrrr}
  \toprule
    & UrlLength& NumNumericChars                      &  ExtMetaScriptLinkRT&SubmitInfoToEmai & ...\\ \midrule
    
    $x$& 112& 15&  0&False& ...\\ %
    
    $x'$& \textbf{113}& \textbf{14}&  0&\textbf{True}& ...\\ \bottomrule
    
  \end{tabular}}
    \caption{\label{table:example-case-study}A
    real-world \textit{phishing} website's feature-space instance, and
    its corresponding perturbed sample created by \sys{}. Perturbed
    features are \textbf{boldfaced} (we omit non-perturbed features).}
\end{table}

\input{tables/exp1-results.tex}

For example, following the \textit{recipe} (\tabref{table:example-case-study}) we
adjusted the website’s URL and HTML. Initially, we were required to
increase the length of the URL by one character (\textit{UrlLength}),
while decreasing a single numerical character from it
(\textit{NumNumericChars}). Secondly, for the HTML, we were required
to introduce the \textit{mailto} function
(\textit{SubmitInfoToEmail}). To avoid affecting other
features and to keep change imperceptible to users, we simply
inserted it in an unreachable function in the JavaScript section.
These minor alterations were sufficient to deceive the detector, while
the website remained functional and visually identical.

\subsection{Feasible Success Rates With \dcs{}}
\label{app:exps:SuccDCs}

In \secref{subsection:exp1-real-success} we evaluated various attacks
on the constrained-feature-space imposed by \dcs{}. We report the
results of this experiment in details in \tabref{table:exp1-results}. 

\subsection{Feasible Success Rates With Valiant's}
\label{appendix:valiants-study}

In \secref{exp-valiants}, we evaluated \sys{} on
the \textit{phishing} dataset while integrating Valiants' constraints to enable
comparison with the work of Sheatsley et
al.~\cite{sheatsley-constraints}.
\tabref{table:exp-sheatsley-extended} reports results on all
datasets.

\begin{table}[t!]
    \centering
  \resizebox{1.0\columnwidth}{!}{%
  \begin{tabular}{ll | rr | rr}
  \toprule

    & & \multicolumn{2}{c|}{Full Attacks} & \multicolumn{2}{c}{Disabling Access} \\
    \textbf{Dataset} & \textbf{Attack}  &   Comp.\ & Comp.\ \& Mis.\ & Comp.\  & Comp.\ \& Mis.\ \\ \midrule

      \multirow{3}{*}{\textit{phishing}} 
      & PGD &  100\% & 67.1\% & 3.3\%	& 3.3\% \\ 
      \cline{2-6} & \tabpgd{} &  100\% & 82.5\% & 54.9\%	& 54.7\% \\ 
      \cline{2-6} & \tabpgd{}+\tabcw{} &  100\% & \textbf{87.6\%} & 77.7\% & 77.7\%  \\ \hline
      
      \hline

    \multirow{3}{*}{\textit{bank}} 
     & PGD &  100\% & 26.9\% & 54.3\%	& 54.3\% \\ 
     \cline{2-6} & \tabpgd{} &  100\% & \textbf{88.5\%}  & 69.9\% & 66.9\% \\ 
     \cline{2-6} & \tabpgd{}+\tabcw{} &  100\% & 87.8\% & 71.8\% & 68.2\% \\ \hline
     
     \hline

      \multirow{3}{*}{\textit{adult}} 
     & PGD &  100\% & 51.2\% & 24.5\%	& 24.5\%  \\ 
     \cline{2-6} & \tabpgd{} &  100\% & 82.7\% & 91.8\%	& 79\%  \\ 
     \cline{2-6} & \tabpgd{}+\tabcw{} &  100\% & \textbf{83.2\%} & 92\% & 79.9\%  \\
    
    \bottomrule
  \end{tabular}}
  \caption{
    \markdiff{Attacking
     with Valiant's constraints, reporting the compliance with the constraints (Comp.) and feasible success rate (Comp. and Mis.) on the full attacks
     and when disabling access to Valiant's constraints.}
    \label{table:exp-sheatsley-extended}
    }
\end{table}

%% file: tables/exp1-results.tex
 \begin{table*}[t!]
 \centering
  \resizebox{0.99\textwidth}{!}{

    \begin{tabular}{lllcccccc}
      \toprule
    
      \bf Model & \bf Dataset  & Attack & C.Acc.? & ${\ell_0}_{\mid M \land C} $ $\downarrow$ & ${\ell_{\infty}}_{\mid M \land C}$ $\downarrow$ & $M$  $\uparrow$ & $C$  $\uparrow$ & $M \land C$ $\uparrow$ \\ \midrule
    
      \multirow{27}{*}{\textit{MLP}} & \multirow{9}{*}{\textit{phishing}}    & TabPGD(1/30, Struc.)  &   $\times$&  5.54 \tiny $\pm 0.04$& 0.007\tiny $\pm 0.0003$& 	99.80\%\tiny $\pm 0.01\%$& 19.43\%\tiny $\pm 0.2\%$& 19.33\%\tiny $\pm 0.3\%$\\  %
      &   & TabPGD(1/30, Struc.) + SAT-Solver(1/30, DCs) & & 5.52 \tiny $\pm 0.06$& 0.009\tiny $\pm 0.0002$&  78.53\%\tiny $\pm 1.2\%$& 88.46\%\tiny $\pm 0.4\%$& \textbf{67.03\%}\tiny $\pm 1.4\%$\\  %
        \cline{3-9}
      
       & & TabPGD(1/30, Struc.) + Tab-CW-L0  &  $\times$& 2.07\tiny $\pm 0.01$& 0.0005\tiny $\pm 0.0003$& 94\%\tiny $\pm 0.6\%$& 49.56\%\tiny $\pm 1.1\%$& \underline{47.43\%}\tiny $\pm 0.9\%$\\ %
       &  & TabPGD(1/30, Struc.) + Tab-CW-L0 + SAT-Solver(1/30, DCs)  & &  3.07\tiny $\pm 0.02$& 0.02\tiny $\pm 0.001$& 63.5\%\tiny $\pm 0.9\%$& 93.3\%\tiny $\pm 1.0\%$& 58.5\%\tiny $\pm 1.4\%$\\ %
      \cline{3-9}
    
      &  & PGD & $\times$ &  - & - & 100\%\tiny $\pm 0\%$& 0.30\%\tiny $\pm 0.2\%$& 0.30\%\tiny $\pm 0.2\%$\\  %
      &  & PGD + SAT-Solver(1/30, DCs) && - & - & 95.8.\%\tiny $\pm 0.3\%$& 6.5\%\tiny $\pm 0.3\%$& 2.3\%\tiny $\pm 0.4\%$\\  %
      \cline{3-9}

      &  & C-PGD(DCs) & &  - & - & 99.36\%\tiny $\pm 0.2\%$& 3.70\%\tiny $\pm 0.7\%$& 3.40\%\tiny $\pm 0.7\%$\\  %
      \cline{3-9}
    
     &   & MoEvA2(Struc.)  & $\times$ & 1.87\tiny $\pm 0.004$& 0.17\tiny $\pm 0.003$& 53.6\%\tiny $\pm 0.1\%$& 79.26\%\tiny $\pm 0.003\%$& 38.66\%\tiny $\pm 0.2\%$\\ %
      &    & MoEvA2(Struc. and DCs)  & &  2.03\tiny $\pm 0.01$& 0.20\tiny $\pm 0.004$& 57.0\%\tiny $\pm 1.0\%$& 80.16\%\tiny $\pm 0.5\%$& 43.46\%\tiny $\pm 1.5\%$\\ %
    \cline{2-9}
     
    \cline{2-9}
      & \multirow{9}{*}{\textit{bank}}    & TabPGD(1/30, Struc.)  &$\times$ &  6.173\tiny $\pm 0.14$& 0.012\tiny $\pm 0.0005$& 	77.10\%\tiny $\pm 1.8\%$& 10.16\%\tiny $\pm 0.4\%$& 9.9\%\tiny $\pm 0.5\%$\\  %
        &  & TabPGD(1/30, Struc.) + SAT-Solver(1/30, DCs) && 6.67\tiny $\pm 0.13$& 0.015\tiny $\pm 0.0006$&  54.46\%\tiny $\pm 2.1\%$& 56.73\%\tiny $\pm 0.2\%$& \textbf{28.43\%}\tiny $\pm 1.2\%$\\  %
        \cline{3-9}
      &  & TabPGD(1/30, Struc.) + Tab-CW-L0 & $\times$ & 2.90\tiny $\pm 0.26$&  0.007\tiny $\pm 0.001$& 78.13\%\tiny $\pm 1.9\%$& 16.2\%\tiny $\pm 0.6\%$& \underline{16.0\%}\tiny $\pm 0.7\%$\\ %
       &   & TabPGD(1/30, Struc.) + Tab-CW-L0 + SAT-Solver(1/30, DCs)  &  &  4.07\tiny $\pm 0.3$& 0.012\tiny $\pm 0.0006$& 44.7\%\tiny $\pm 2.3\%$& 66.36\%\tiny $\pm 0.2\%$& \textbf{27.26\%}\tiny $\pm 1.3\%$\\ %
      \cline{3-9}
    
      &  & PGD &$\times$&  - & - & 100.0\%\tiny $\pm 0.0\%$& 4.93\%\tiny $\pm 0.4\%$& 4.93\%\tiny $\pm 0.4\%$\\  %
       &   & PGD + SAT-Solver(1/30, DCs) & & 7.85\tiny $\pm 0.18$& 0.16\tiny $\pm 0.0001$& 30.9\%\tiny $\pm 1.2\%$& 48.46\%\tiny $\pm 2.0\%$& 14.1\%\tiny $\pm 0.9\%$\\  %
      \cline{3-9}

      &  & C-PGD(DCs) &&  - & - & 100.0\%\tiny $\pm 0.0\%$& 10.9\%\tiny $\pm 0.4\%$&10.9\%\tiny $\pm 0.4\%$\\  %
      \cline{3-9}
    
     &   & MoEvA2(Struc.)  &$\times$& 1.29\tiny $\pm 0.1$& 0.006\tiny $\pm 0.0008$& 20.73\%\tiny $\pm 0.5\%$& 66.40\%\tiny $\pm 0.7\%$& 8.43\%\tiny $\pm 0.05\%$\\
      &    & MoEvA2(Struc. and DCs)  &&  2.78\tiny $\pm 0.04$& 0.07\tiny $\pm 0.01$& 26.73\%\tiny $\pm 1.2\%$& 47.5\%\tiny $\pm 0.4\%$& 8.8\%\tiny $\pm 0.05\%$\\
    \cline{2-9}

    \cline{2-9}
     &  \multirow{9}{*}{\textit{adult}}    & TabPGD(1/30, Struc.)  &$\times$&  4.44\tiny$\pm 0.13$& 0.011\tiny$\pm 0.0008$& 86.63\%\tiny$\pm 1.2\%$& 16.7\%\tiny$\pm 0.3\%$& 16.3\%\tiny$\pm 0.2\%$\\  %
     &     & TabPGD(1/30, Struc.) + SAT-Solver(1/30, DCs)  && 5.09\tiny$\pm 0.09$& 0.015\tiny$\pm 0.0006$&  72.63\%\tiny$\pm 1.0\%$& 54.03\%\tiny$\pm 0.8\%$& \textbf{32.5\%}\tiny$\pm 0.8\%$\\  %
        \cline{3-9}
    &    & TabPGD(1/30, Struc.) + Tab-CW-L0  & $\times$ & 2.32\tiny $\pm 0.18$& 0.007\tiny $\pm 0.007$& 87.53\%\tiny $\pm 1.5\%$& 20.8\%\tiny $\pm 0.5\%$& \underline{20.36\%}\tiny $\pm 0.6\%$\\ %
     &     & TabPGD(1/30, Struc.) + Tab-CW-L0 + SAT-Solver(1/30, DCs) & &  3.19\tiny $\pm 0.06$& 0.011\tiny $\pm 0.0003$& 65.16\%\tiny $\pm 2.4\%$& 61.7\%\tiny $\pm 1.7\%$& \textbf{32.4\%}\tiny $\pm 1.3\%$\\ %
      \cline{3-9}
    
      &  & PGD &$\times$&  - & - & 100.0\%\tiny $\pm 0.0\%$& 3.50\%\tiny $\pm 0.1\%$& 3.50\%\tiny $\pm 0.1\%$\\  %
       &   & PGD + SAT-Solver(1/30, DCs) && 9.71\tiny $\pm 0.54$& 0.01\tiny $\pm 0.002$& 65.06\%\tiny $\pm 0.007\%$& 21.16\%\tiny $\pm 0.008\%$& 8.50\%\tiny $\pm 1.4\%$\\  %
      \cline{3-9}

      &  & C-PGD(DCs) &&  - & - & 100.0\%\tiny $\pm 0.0\%$& 25.53\%\tiny $\pm 2.8\%$&25.36\%\tiny $\pm 2.8\%$\\  %
      \cline{3-9}
    
      &  & MoEvA2(Struc.)  &$\times$ & 3.30\tiny $\pm 0.24$& 0.03\tiny $\pm 0.003$& 52.96\%\tiny $\pm 0.7\%$& 49.33\%\tiny $\pm 0.8\%$& \underline
{18.4\%}\tiny $\pm 0.5\%$\\
      &    & MoEvA2(Struc. and DCs)  &&  5.95\tiny $\pm 0.23$& 0.09\tiny $\pm 0.004$& 58.10\%\tiny $\pm 0.9\%$& 47.06\%\tiny $\pm 0.6\%$& 20.5\%\tiny $\pm 1.5\%$\\
    \hline

    \hline
      \multirow{12}{*}{\textit{TabNet}} & \multirow{4}{*}{\textit{adult}}    
           & TabPGD(1/30, Struc.) + Tab-CW-L0  & $\times$ & 1.35& 0.001& 82.0\%& 22.2\% & 22.0\% \\ 
       &   & TabPGD(1/30, Struc.) + Tab-CW-L0 + SAT-Solver(1/30, DCs) & & 3.81 & 0.011 & 71.5\% & 60.0\% & \textbf{43.7\%} \\  
      \cline{3-9}
    
      &  & MoEvA2(Struc.)  &$\times$ & 1.92& 0.048 & 56.9\% & 58.5\% & 27.2\%  \\
      &  & MoEvA2(Struc. and DCs)  &&   3.74 & 0.088 & 61.2\% & 57.9\% & 30.7\% \\
    \cline{2-9}

       & \multirow{4}{*}{\textit{phishing}}    
           & TabPGD(1/30, Struc.) + Tab-CW-L0  & $\times$ & 3.28& 0.01& 83.7\%& 32.6\% & 30.0\% \\ 
       &   & TabPGD(1/30, Struc.) + Tab-CW-L0 + SAT-Solver(1/30, DCs) & &  3.7 & 0.05 & 46.1\% & 93.6\% & \textbf{44.1\%} \\  
      \cline{3-9}
    
      &  & MoEvA2(Struc.)  &$\times$ & 1.90 & 0.16 & 55.7\% & 71.5\% & 33.5\%  \\
      &  & MoEvA2(Struc. and DCs)  &&  1.99 & 0.22 & 52.7\% & 79.0\% & 37.3\% \\
    \cline{2-9}

       & \multirow{4}{*}{\textit{bank}}    
           & TabPGD(1/30, Struc.) + Tab-CW-L0  & $\times$ & 2.52& 0.003 & 52.5\%& 10.0\% & 9.2\% \\ 
       &   & TabPGD(1/30, Struc.) + Tab-CW-L0 + SAT-Solver(1/30, DCs) & & 4.58 & 0.009 & 48.0\% & 42.6\% & \textbf{17.6\%} \\  
      \cline{3-9}
    
      &  & MoEvA2(Struc.)  &$\times$ & 1.16 & 0.027 & 21.0\% & 66.8\% & 7.8\%  \\
      &  & MoEvA2(Struc. and DCs)  & & 2.12 & 0.193 & 27.8\% & 62.3\% & 12.6\% \\
    \bottomrule

  \end{tabular}}
  \caption{\label{table:exp1-results}Evaluation of our adversarial tabular attack variants (\cafa{}), compared to other attacks. For each attack, we mention the used parameters/information in parentheses by $AttackName(\epsilon, Constraints)$ (where $\epsilon$ stands for the \textit{standardized-}$\ell_\infty$, and \textit{constraints} names the type of constraints incorporated). We also state whether the attack used access to constraints (C. Acc.?) Then, we compare the $\ell_0$ difference from the original sample ($\ell_0$) averaged over the \textit{realizable and successfully mislcassified} adversarial samples, the average \textit{standardized-} $\ell_{\infty}$ on the same subset ($l_{\infty}$), the miss-classification rate ($M$), the compliance with the \dcs{} ($C$) and finally the combination of the latter two, which means the rate of realizable adversarial samples ($M \land S$).
The best-performing attack, for each dataset and model, is marked with \textbf{bold}, and under the setting with \textit{no} access to constraints we mark the best-performing attack with \underline{underline}. }
\end{table*}

%% file: meta-review.tex
\clearpage %

\section{Meta-Review}

The following meta-review was prepared by the program committee for the 2024
IEEE Symposium on Security and Privacy (S\&P) as part of the review process as
detailed in the call for papers.

\subsection{Summary}

In this paper, the authors investigate the vulnerability of machine learning models in domains with constraints. The authors propose an approach built from a series of algorithms for robustness evaluations, namely, mining constraints, enforcing datatype constraints during crafting, and projecting examples to comply with the learned domain constraints. Specifically, the authors propose using denial constraints to mine constraints from examples, integrating datatype constraints during the adversarial crafting process, and using SAT solvers (when necessary) to project invalid examples onto a constraint-complaint space. In their evaluation, the authors investigate robustness evaluations with 3 datasets, against existing baselines, and observe various performance curves of constraint-learning configurations showing constraint-complaint adversarial examples can be often found.

\subsection{Scientific Contributions}

\begin{itemize}
\item Independent Confirmation of Important Results with Limited Prior Research.
\item Creates a New Tool to Enable Future Science.
\item Provides a Valuable Step Forward in an Established Field.
\end{itemize}

\subsection{Reasons for Acceptance}
\begin{enumerate}
\item This paper provides a valuable step forward in an established field. Specifically, the authors develop \sys{} to craft adversarial examples while satisfying mined constraints.
\item Experimental results demonstrate that \sys{} achieves a higher (feasible) attack success rate with lower attack costs compared to existing attacks, including \pgd, C-PGD, and MoEvA2.
\end{enumerate}